\documentclass[11pt]{article} 
\usepackage[utf8]{inputenc}
\usepackage{multirow}
\usepackage{amsfonts}
\usepackage{amsmath}
\usepackage{amssymb}
\usepackage{amsthm}
\usepackage{mathrsfs}
\usepackage{comment}
\usepackage{caption}
\usepackage{color}
\usepackage[dvipsnames]{xcolor}
    \definecolor{darkgreen}{rgb}{0,0.5,0}
    \definecolor{darkblue}{rgb}{0,0,0.6}
    \definecolor{purple}{rgb}{0.4,.2,0.7}
\usepackage[margin = 2.7cm]{geometry}
\usepackage{graphicx}
\usepackage[hyperfootnotes = false, colorlinks = true, linkcolor = darkblue, citecolor = purple]{hyperref}
\usepackage{subcaption}
\usepackage{microtype}
\usepackage{amsmath,braket}
\usepackage[toc,page]{appendix}
\usepackage{tikz}
\numberwithin{figure}{section}
\numberwithin{equation}{section}
\numberwithin{table}{section}

\newcommand{\de}{\partial}
\newcommand{\be}{\begin{equation}}
\newcommand{\ba}{\begin{eqnarray}}
\newcommand{\ea}{\end{eqnarray}}
\newcommand{\ee}{\end{equation}}

\newcommand{\f}{\frac}

\newcommand{\s}{\sqrt}
\newcommand{\vp}{\varphi}

\newcommand{\ap}{\alpha}

\newcommand{\ddd}{\cdot\cdot\cdot}
\newcommand{\no}{\nonumber \\}

\newcommand{\bea}{\begin{eqnarray}}
\newcommand{\eea}{\end{eqnarray}}
\newcommand{\bes}{\begin{equation*}}
\newcommand{\beas}{\begin{eqnarray*}}
\newcommand{\eeas}{\end{eqnarray*}}
\newcommand{\bas}{\begin{array*}}
\newcommand{\eas}{\end{array*}}
\newcommand{\ees}{\end{equation*}}
\newcommand{\nn}{\nonumber}

\newcommand{\ep}{\epsilon}

\makeatother
\usepackage{comment}
\usepackage{babel}
\newcommand{\mpl}{M_{\rm pl}}
\allowdisplaybreaks

\hypersetup{
  colorlinks=true,
  linkcolor=darkblue,
  citecolor=purple,
  urlcolor=darkblue
}

\usepackage{mathtools}

\begin{document}

	\begin{flushright}
		YITP-26-125
        \\
         RIKEN-iTHEMS-Report-26
	\end{flushright}

\thispagestyle{empty}
\begin{center}
    ~\vspace{3mm}

 {\fontsize{20}{20}\selectfont\bfseries 
 Effective Dynamics of Inflationary End-of-the-World Branes in AdS$_3$
 \\}
     
   \vspace{0.5in}
     
   {\large \bf Kosei Fujiki$^1$,  Michitaka Kohara$^1$, Jason Kristiano$^{1,4}$, \\
   \vspace{3mm}
   Yu-ki Suzuki$^2$ and Tadashi Takayanagi$^{1,3}$
   }

    \vspace{0.5in}
 
   ~
   \\
  {\fontsize{11}{13}\selectfont
$^1$ \textit{Center for Gravitational Physics and Quantum Information,\\
Yukawa Institute for Theoretical Physics, \\Kyoto University, Kyoto 606-8502, Japan}\\
 \vspace{3mm}
$^2$ \textit{RIKEN Center for Interdisciplinary Theoretical
and Mathematical Sciences (iTHEMS),
Wako, Saitama 351-0198, Japan}\\
 \vspace{3mm}
$^3$ \textit{Inamori Research Institute for Science,\\
	620 Suiginya-cho, Shimogyo-ku, Kyoto 600-8411 Japan}\\
\vspace{3mm}
$^4$ \textit{Research Center for the Early Universe (RESCEU),\\
Graduate School of Science, The University of Tokyo, Tokyo 113-0033, Japan}
}

    \vspace{0.3in}

\end{center}

\vspace{0.3in}

\begin{abstract}
 We develop an effective description of a two-dimensional cosmological end-of-the-world brane embedded in AdS$_3$, with a scalar field localized on the brane. Integrating out bulk degrees of freedom, especially in the small-gradient regime, the effective action reduces to a Liouville-like theory coupled to the brane scalar. We then construct brane trajectories whose induced geometry realizes slow-roll inflation and reconstruct the associated scalar profile and potential. We also study regular Euclidean brane geometries that admit a smooth continuation to Lorentzian de Sitter and inflationary universes, and evaluate the semiclassical on-shell action for the de Sitter solution. Finally, we analyze linearized inhomogeneous perturbations and find no exponentially growing mode within the regime of validity of the approximation.
 \end{abstract}

\vspace{1in}

\pagebreak

\renewcommand{\contentsname}{\texorpdfstring{\textcolor{darkblue}{Contents}}{Contents}}
{\large
\tableofcontents
}

\clearpage
\section{Introduction and summary}

A major direction in gravitational physics is to explore theories that extend Einstein gravity and may capture physics beyond its regime of validity. Among the various approaches, one particularly useful framework is to regard our spacetime as a lower-dimensional brane embedded in a higher-dimensional gravitational theory \cite{Randall:1999ee,Randall:1999vf}. In such braneworld scenarios, Einstein gravity can be recovered at sufficiently low energies, while deviations from it arise at higher energies as a consequence of the dynamics of the higher-dimensional bulk \cite{Garriga:1999yh,Shiromizu:1999wj}. From the lower-dimensional point of view, these corrections encode the backreaction of the bulk geometry on the brane and therefore provide a concrete setting in which modifications of Einstein gravity can be studied within a geometrical framework.


Several features make the braneworld scenario particularly promising. In warped backgrounds, gravitational degrees of freedom can be localized near a brane, so that lower-dimensional gravity emerges over an appropriate range of scales even though the fundamental gravitational theory lives in a higher-dimensional spacetime. From the brane perspective, deviations from Einstein gravity can then be understood geometrically as effects associated with the embedding and with gravitational degrees of freedom in the bulk. This picture also admits a natural holographic interpretation when the bulk is asymptotically AdS: the higher-dimensional gravitational dynamics can be related to quantum degrees of freedom coupled to gravity on the brane \cite{Gubser:1999vj,Karch:2000ct,Shiromizu:2001jm,Koyama:2001rf,Shiromizu:2001ve}.
We can view this as a generalization of the AdS/CFT correspondence \cite{Maldacena:1997re,Gubser:1998bc,Witten:1998qj}. Moreover, when the brane tension is enough small, the brane world-volume becomes asymptotically AdS and thus we can employ the AdS/CFT duality once again. This leads to the AdS/BCFT duality \cite{Takayanagi:2011zk,Fujita:2011fp} (see also \cite{Karch:2000ct}), which argues that the gravitational theory in region between the brane and the asymptotic AdS boundary is dual to a conformal field theory on a space with boundaries, i.e. a boundary conformal field theory (BCFT) \cite{Cardy:1984bb,Cardy:2004hm}.  
A closely related generalization is the AdS/ICFT correspondence \cite{Bachas:2001vj,Anous:2022wqh,Liu:2024oxg}, which extends the holographic framework to conformal field theories with interfaces. In this correspondence, an interface in the boundary CFT is described holographically by a corresponding interface or defect in the asymptotically AdS bulk, and the resulting gravitational theory is dual to an interface conformal field theory (ICFT).

Cosmology provides an important field where we can apply the braneworld perspective. Despite substantial progress, obtaining controlled top-down realizations of de Sitter space and slow-roll inflation in quantum gravity remains a major challenge. In string compactifications, for example, stabilizing the compactification moduli and maintaining control over quantum and higher-derivative corrections must be reconciled with positive vacuum energy and a sufficiently flat inflaton potential. These difficulties motivate a complementary approach based on braneworld holography, in which cosmological spacetimes arise as induced geometries on EOW branes embedded in AdS, with lower-dimensional quantum gravity effects expected to be encoded in the higher-dimensional bulk description. The motion and deformation of a brane in a higher-dimensional spacetime can induce a time-dependent geometry on its worldvolume, allowing the brane itself to be interpreted as an evolving universe. In the AdS/BCFT setup \cite{Takayanagi:2011zk,Fujita:2011fp}, this idea has already led to simple models of cosmological evolution, including de Sitter and big-bang-like geometries realized as induced spacetimes on an end-of-the-world brane\cite{Cooper:2018cmb,Antonini:2019qkt,VanRaamsdonk:2021qgv,Omiya:2021olc,Antonini:2022blk,Waddell:2022fbn,Fujiki:2025yyf}.
From the viewpoint of brane-world holography, the classical dynamics of the higher-dimensional bulk can then encode lower-dimensional quantum-gravitational cosmology on the brane. Of particular interest is the possibility of describing inflation in this framework. Inflation corresponds to an epoch of accelerated expansion that is well approximated by de Sitter spacetime when the Hubble parameter $H$ is nearly constant. In conventional slow-roll inflation, a scalar field evolves sufficiently slowly along its potential so that the evolution deviates only weakly from exact de Sitter space. This deviation can be characterized by the slow-roll parameter $\epsilon_1 \equiv - \dot{H}/H^2$, with $\epsilon_1 \ll 1$ during the quasi-de Sitter regime and the end of inflation associated with the breakdown of the slow-roll approximation. Such a controlled deviation from de Sitter space provides a natural target for cosmological brane models: rather than considering only an exactly de Sitter brane, one would like to understand whether the brane dynamics can reproduce a slowly evolving inflationary geometry together with the matter configuration responsible for it.

Another important approach to holography in cosmological spacetimes is the dS/CFT correspondence, which proposes that quantum gravity in asymptotically de Sitter space can be described in terms of a Euclidean conformal field theory living at its asymptotic boundary \cite{Strominger:2001pn,Witten:2001kn,Spradlin:2001pw,Bousso:2001mw,Balasubramanian:2002zh,Klemm:2001ea}. In this framework, cosmological evolution is closely related to scale or renormalization-group evolution in the boundary theory, and bulk fields are associated with boundary operators with characteristic conformal dimensions \cite{Balasubramanian:2001nb}. The dS/CFT correspondence has been applied to various problems in cosmology \cite{Strominger:2001gp,Maldacena:2002vr,Larsen:2002et,Larsen:2003pf,McFadden:2009fg,McFadden:2010na}.
In this work, we realize a lower-dimensional dS/CFT correspondence within the AdS/BCFT framework, relating a two-dimensional de Sitter brane to the one-dimensional boundary of the BCFT.

\begin{figure}[htbp]
		\centering
		\includegraphics[width=5cm]{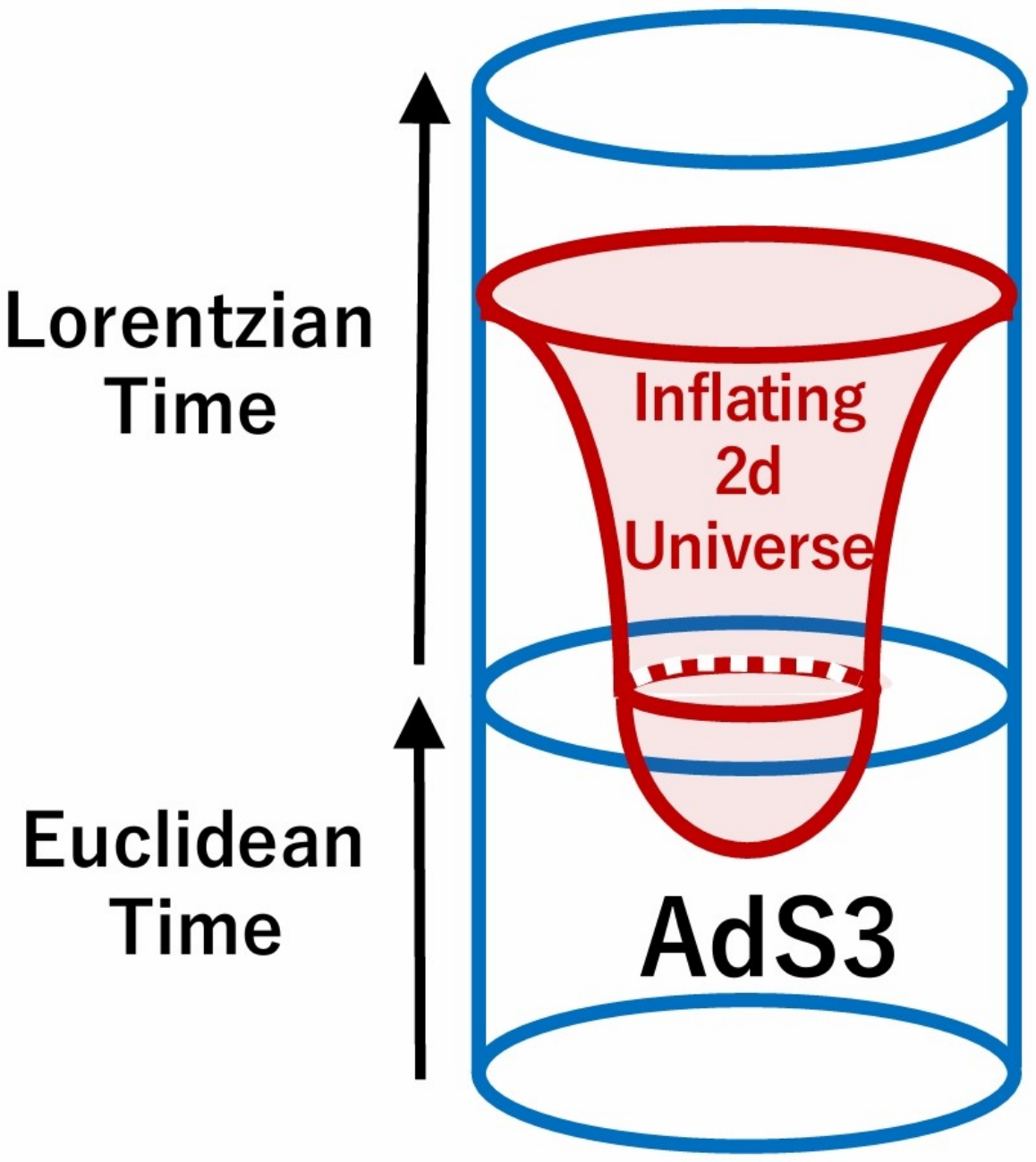}
		\caption{A sketch of our holographic description of two-dimensional inflating Universe embedded in AdS$_3$. The quantum cosmology in the two-dimensional inflating universe on the brane, depicted by the red surface, is dual to the classical gravity in the three-dimensional bulk region which is inside the brane.}
		\label{fig:inflationBCFT}
\end{figure}

In the present work, we realize such a slow-roll geometry directly as the induced metric on an EOW brane in AdS,  as sketched in Fig.~\ref{fig:inflationBCFT}. We can employ a de Sitter brane in AdS, which is expected to be dual to a BCFT with a spacelike boundary \cite{Akal:2020wfl,Chen:2020tes} (see \cite{Wei:2024zez,Fujiki:2025yyf,Hoback:2026yqj} for important updates). Then we introduce a scalar field on the brane which is identified with the inflaton, dual to the boundary RG flow via the AdS/BCFT \cite{Kanda:2023zse, Kanda:2023jyi, Fujiki:2025yyf,Maeda:2026awj}. Starting from the desired inflationary scale factor, we determine the corresponding brane trajectory and reconstruct the profile and potential of the scalar field localized on the brane. The deformation away from the de Sitter brane is controlled by the slow-roll parameter, and the resulting scalar potential exhibits the behavior expected for slow-roll evolution.   This gives a concrete setting in which inflationary dynamics can be studied simultaneously from the viewpoints of the higher-dimensional brane embedding and the lower-dimensional effective gravitational theory.

At the same time, this higher-dimensional description introduces an important difficulty. The bulk gravitational degrees of freedom can propagate away from the brane and subsequently influence its dynamics. Consequently, from the brane perspective, the bulk backreaction generally appears as a nonlocal interaction. The effective gravitational equations on the brane are therefore not closed in terms of brane variables alone: their evolution depends on information about the bulk geometry. Despite substantial progress in understanding braneworld dynamics, a general local effective action that fully incorporates these nonlocal bulk effects has not been established. Four-dimensional effective action on the brane in AdS$_5$ was discussed in \cite{Kanno:2002iaa}.


In this work, we obtain a partial but analytical answer to this problem by restricting to the cosmological brane in a three-dimensional bulk AdS. As is well known, three-dimensional Einstein gravity possesses no local propagating gravitational degrees of freedom. This special property considerably simplifies the braneworld dynamics: there are no independent local bulk gravitational modes that must be evolved in addition to the brane degrees of freedom, and the dynamics can consequently be described by equations that close on the brane. Exploiting this simplification, we integrate out the bulk gravitational variables while retaining the dynamical degree of freedom associated with the brane embedding. In this way, we derive an effective action governing the motion of the brane. Refer to \cite{Suzuki:2022xwv,Geng:2022slq,Neuenfeld:2024gta} for earlier works on the effective action for the AdS brane.

We further investigate the relation between this effective theory and familiar theories of two-dimensional gravity. In particular, after integrating out the AdS$_3$ bulk we obtain an exact gauge-fixed action for the brane embedding profile and the scalar field $\phi$ localized on the brane. In the regime in which gradients of the brane profile are small, it is convenient to introduce a Liouville-like field $\Phi$ from the Weyl factor of its induced metric. Expanding the exact action to second order in derivatives and a small-gradient approximation, where the brane is at asymptotic region,  then yields a Liouville-like theory coupled to the brane scalar field. 
Related effective actions for fluctuating finite-cutoff surfaces in \(\mathrm{AdS}_3\) have been studied in Refs.~\cite{Callebaut:2025thw,Callebaut:2026hso}. In particular, Ref.~\cite{Callebaut:2025thw} derived a finite-cutoff generalization of the Liouville action in Poincaré AdS.
A subtlety, however, is that the gauge-fixed action does not retain all the information contained in the original brane equations of motion. We carefully study this and 
after restoring a covariant description, we reproduce the lost constraints and confirm that it is equivalent to the brane EOM.
This provides an analytical description of the cosmological brane dynamics and makes it possible to formulate the system entirely in terms of degrees of freedom intrinsic to, or associated with the embedding of, the brane.

An additional advantage of having an explicit effective action is that it allows us to study the wave function of the universe semiclassically for various brane configurations. Based on the seminal framework for defining such a wave function developed by Hartle and Hawking  \cite{Hartle:1983ai} in the context of quantum cosmology, the basic idea is to characterize quantum gravity in terms of transition amplitudes between spatial geometries. Since a generally covariant formulation does not single out an external time parameter, such an amplitude is naturally represented by a gravitational path integral over spacetime geometries whose boundaries are the prescribed spatial geometries.

When this framework is applied to the origin of the universe, however, specifying an initial spatial geometry of finite size amounts to imposing an additional initial condition on the cosmological state. The Hartle--Hawking no-boundary proposal provides a prescription for avoiding such an arbitrary initial boundary condition. Rather than assuming an initial spatial boundary, the gravitational path integral is taken over geometries that close off smoothly in the past. The initial condition is thus replaced by a regularity condition on the geometry itself. At the semiclassical level, the corresponding wave function is determined by the on-shell action of regular saddle-point geometries satisfying the prescribed final boundary condition. This perspective is also closely related to dS/CFT, where the late-time wave function of an asymptotically de Sitter universe can be interpreted holographically in terms of a partition function of a boundary field theory \cite{Maldacena:2002vr,Harlow:2011ke,Hertog:2011ky}. We also present an analytical solution which describes a class of inflation processes.

There remain, of course, fundamental questions concerning this prescription. These include whether the resulting wave function can be consistently normalized and endowed with a probabilistic interpretation, as well as the more basic issue of how the gravitational path integral and its integration contour should be defined. We do not attempt to resolve these questions in the present work. Instead, taking the semiclassical gravitational path integral as our starting point, we evaluate the corresponding on-shell weight,
\[
\Psi \sim e^{-I_{\mathrm{on\text{-}shell}}},
\]
for de Sitter and inflationary geometries realized on the brane. Even though we perform this analysis in our lower dimensional setups, we can calculate the value of wave function analytically because our brane model with a localized scalar is exactly solvable.   
Even though this analysis using the brane is analyzed at the semi-classical level, the result is expected to include quantum corrections owing the braneworld holography. Therefore, this allows us to compare different brane profiles and determine which configurations are favored at the semiclassical level. 

Finally, we analyze the stability of the slow-roll brane solution against small inhomogeneous perturbations. We perturb both the brane embedding and the localized scalar field and linearize the Neumann boundary condition around the homogeneous cosmological background. The scalar perturbation is constrained by the deformation of the brane embedding, leaving an equation for the physical geometric perturbation. For the slow-roll solution considered here, its Fourier modes obey an oscillatory wave equation and exhibit no exponentially growing mode. The inflationary brane is therefore linearly stable against these perturbations within the regime of validity of our approximation.

This paper is organized as follows. In section two, we will discuss the brane description of the two-dimensional de Sitter space embedded in three-dimensional AdS space with a localized scalar on the brane. We will analyze its effective action on the brane and work out the full constraint equations which are missing in the naive construction of effective action. By assuming the deviation from the de Sitter space is small, we will derive the covariantized effective action and show that it describes the dynamics of cosmological brane perfectly without imposing any constraints. In section three, starting from a brief review of inflation, we will calculate slow roll solutions on the dS brane embedded into Poincare AdS$_3$. In section four, we will analyze the Hartle-Hawking initial condition of the brane solution. We find a class of analytical brane solutions which describes the slow roll inflation. We also compute the on-shell action which is related to the probability of the nucleation of dS brane. In section five, we will analyze the small perturbations of our brane solution to confirm its stability. In appendix A, we present the details of our minisuperspace reduction. In appendix B, we show detailed calculations of extrinsic curvatures.

\section{Brane model} \label{sec:BraneModel}

\subsection{Basic description of the model}
We consider the AdS$_3$/BCFT$_2$ setup, in which an end-of-the-world (EOW) brane terminates the asymptotically AdS$_3$ spacetime and is dual to a boundary of the BCFT$_2$~\cite{Takayanagi:2011zk,Fujita:2011fp}. Following Refs.~\cite{Kanda:2023zse,Fujiki:2025yyf}, we introduce a scalar field localized on the EOW brane in order to allow for nontrivial brane dynamics.
We assume the following action in this spacetime:
\begin{equation}
\begin{aligned}
\label{eq:BraneScalarLagrangian}
I=&\frac{1}{16\pi G_N}\int_{M} d^3x \s{-g}(R-2\Lambda)+\frac{1}{8\pi G_N}\int_{N}d^2x\s{-h}K  \\
&+\frac{1}{8\pi G_N}\int_{Q} d^2x\s{-h}\left(K-h^{ab}\de_a\phi\de_b\phi-V(\phi)\right),
\end{aligned}
\end{equation}
where the cosmological constant is fixed as $\Lambda=-1$. $M$ denotes the three-dimensional bulk and $N$ is the asymptotic boundary. $Q$ is the EOW brane. The variation of this action gives
\ba
\delta I=\frac{1}{16\pi G_N}\int_{Q} d^2x \s{-h}\delta h^{ab}\left[
K_{ab}-h_{ab}K-2\de_a\phi\de_b\phi+h_{ab}\left(h^{cd}\de_c\phi\de_d\phi+V(\phi)\right)\right],
\ea
where $h_{ab}$ is the induced metric and $K_{ab}$ is the extrinsic curvature. Its trace is denoted by $K=h^{ab}K_{ab}$. Now, we impose the Neumann boundary condition on the EOW brane, which is written by
\begin{align}
&K_{ab}-h_{ab}K=T_{Q\,ab},\quad T_{Qab}\equiv 2\de_a\phi\de_b\phi-h_{ab}\left(h^{cd}\de_c\phi\de_d\phi+V(\phi)\right).  
\label{eq:generalNBC}
\end{align}
This condition relates the dynamics of the EOW brane to the behavior of the scalar field.
The equation of motion for the localized scalar field is obtained as follows:
\begin{equation}
    2\de_a(\s{-h}h^{ab}\de_b\phi)-\s{-h}V'(\phi)=0.
\end{equation}
Recalling that $\nabla^a(K_{ab}-h_{ab}K)=0$ is satisfied in three-dimensional AdS spacetime and taking \eqref{eq:generalNBC} into account, we find that this equation of motion follows automatically. 

Let us fix the background metric to be the Poincar\'e coordinates:
\begin{align}
    ds^2=\frac{dz^2-dt^2+dx^2}{z^2}.  \label{PAdS}
\end{align}
We insert the EOW brane whose trajectory is given by $z=Z(t)$. Here, we assume the translational symmetry along the $x$ direction for simplicity. There are two possible choices for how to choose the bulk region in this setup:
\ba
\mbox{Type I:}\ \ \ 0\leq z\leq Z(t),\quad \mbox{Type II:}\ \ \ z\geq Z(t).
\ea
Then, the induced metric is written by
\begin{align}
    ds^2=\frac{-(1-\dot{Z}^2)dt^2+dx^2}{Z^2},
    \label{eq:Poincareinduced}
\end{align}
where $\dot{Z}=\f{dZ}{dt}$. We also impose the timelike condition $1-\dot{Z}^2>0$. In this setup, the Neumann boundary conditions can be rewritten as
\begin{align}
    \dot{\phi}^2&=-\frac{\epsilon\ddot{Z}}{2Z\s{1-\dot{Z}^2}}, \label{phid2_Z}\\
    V&=\frac{\epsilon Z\ddot{Z}}{2(1-\dot{Z}^2)^{3/2}}-\frac{\epsilon}{\sqrt{1-\dot{Z}^2}}. \label{V_Z}
\end{align}
Here, we have introduced $\epsilon=\pm1$ to distinguish between Type I and Type II. We note that $\epsilon=+1$ and $\epsilon=-1$ correspond to Type I and Type II, respectively. From these conditions, we immediately obtain the energy-conservation-like relation:
\begin{equation}
    \f{Z^2}{\sqrt{1-\dot{Z}^2}}\dot{\phi}^2+\sqrt{1-\dot{Z}^2}V(\phi)=-\epsilon.
\end{equation}
In the following parts of this section, we focus on the Type II case. We note that calculations in the Type I case are almost the same.

 Notice that if we recover the AdS radius, the dS radius on the brane is represented as
 \be
R_{\rm dS}=\frac{R_{\rm AdS}}{\sqrt{V^2-1}},
 \ee
where we consider the straight dS brane.

\subsection{Gauge-unfixed minisuperspace reduction}
\label{sec:gauge-unfixed-minisuperspace}

The homogeneous ansatz is often introduced directly in the static gauge
$t=\sigma$.  Such a procedure is adequate for deriving the local evolution
equations, but it obscures the equation obtained by varying the embedding time.
To retain the complete constraint structure, we first perform the
minisuperspace reduction without fixing the worldvolume time coordinate.  In
this subsection we take the Type II orientation, $\epsilon=-1$, as stated
above.
\subsubsection*{Minisuperspace ansatz}
We parametrize the EOW brane by $(\sigma,x)$ and consider
\begin{align}
 X^\mu(\sigma,x)&=\bigl(t(\sigma),Z(\sigma),x\bigr),
 &\phi&=\phi(\sigma),
 &t'^{\,2}-Z'^{\,2}&>0,
 \label{eq:unfixed-embedding}
\end{align}
where a prime denotes a derivative with respect to $\sigma$.  The induced
metric, its inverse, and its determinant are given by
\begin{align}
 h_{\sigma\sigma}&=-\frac{t'^{\,2}-Z'^{\,2}}{Z^2},
 &h_{xx}&=\frac{1}{Z^2},
 &h_{\sigma x}&=0,
 \label{eq:unfixed-induced-metric}\\
 h^{\sigma\sigma}&=-\frac{Z^2}{t'^{\,2}-Z'^{\,2}},
 &h^{xx}&=Z^2,
 &\sqrt{-h}&=\frac{\sqrt{t'^{\,2}-Z'^{\,2}}}{Z^2}.
 \label{eq:unfixed-induced-inverse}
\end{align}
For the normal orientation appropriate to Type II, the trace of the extrinsic
curvature is
\begin{equation}
 K=
 \frac{-Zt'Z''+Zt''Z'+2t'\bigl(t'^{\,2}-Z'^{\,2}\bigr)}
 {\bigl(t'^{\,2}-Z'^{\,2}\bigr)^{3/2}}.
 \label{eq:unfixed-K}
\end{equation}
The detailed computation of \eqref{eq:unfixed-K} is given in
Appendix~\ref{app:unfixed-reduction-details}. 
For the type I, we just flip the sign of $\ep$ and obtain the action with additional $(-1)$ multiplied.

After integrating over the translationally invariant direction, the brane
contribution becomes
\begin{align}
 I_{Q}=\frac{L_x}{8\pi G_N}\int _Qd\sigma\,
 \left[
 \frac{2t'}{Z^2}
 +\frac{t''Z'-t'Z''}{Z\bigl(t'^{\,2}-Z'^{\,2}\bigr)}
 +\frac{\phi'^{\,2}}{\sqrt{t'^{\,2}-Z'^{\,2}}}
 -\frac{\sqrt{t'^{\,2}-Z'^{\,2}}}{Z^2}V(\phi)
 \right].
 \label{eq:unfixed-brane-action}
\end{align}
Up to terms independent of the brane trajectory, the renormalized on-shell
bulk contribution for the Type II region is
\begin{equation}
 I_{\mathrm{bulk}}^{\mathrm{on\text{-}shell}
 }=-\frac{L_x}{8\pi G_N}\int d\sigma\,\frac{t'}{Z^2}.
 \label{eq:unfixed-bulk-action}
\end{equation}
Consequently, the exact gauge-unfixed minisuperspace action is
\begin{align}
 I_{\mathrm{mini}}
 =\frac{L_x}{8\pi G_N}\int d\sigma\,
 \left[
 \frac{t'}{Z^2}
 +\frac{t''Z'-t'Z''}{Z\bigl(t'^{\,2}-Z'^{\,2}\bigr)}
 +\frac{\phi'^{\,2}}{\sqrt{t'^{\,2}-Z'^{\,2}}}
 -\frac{\sqrt{t'^{\,2}-Z'^{\,2}}}{Z^2}V(\phi)
 \right]
 +I_{\mathrm{corner}}.
 \label{eq:unfixed-mini-second-derivative}
\end{align}
No gauge condition has been imposed in this expression.

The second-derivative term can be isolated by using
\begin{equation}
 \frac{t''Z'-t'Z''}{Z\bigl(t'^{\,2}-Z'^{\,2}\bigr)}
 =-\frac{1}{Z}\frac{d}{d\sigma}
 \operatorname{arctanh}\!\left(\frac{Z'}{t'}\right).
 \label{eq:unfixed-arctanh-identity}
\end{equation}
After an integration by parts, one obtains
\begin{align}
 I_{\mathrm{mini}}
 =&\frac{L_x}{8\pi G_N}\int d\sigma\,
 \left[
 \frac{t'}{Z^2}
 -\frac{Z'}{Z^2}\operatorname{arctanh}\!\left(\frac{Z'}{t'}\right)
 +\frac{\phi'^{\,2}}{\sqrt{t'^{\,2}-Z'^{\,2}}}
 -\frac{\sqrt{t'^{\,2}-Z'^{\,2}}}{Z^2}V(\phi)
 \right]
 \nonumber\\
 &-\frac{L_x}{8\pi G_N}
 \left[
 \frac{1}{Z}\operatorname{arctanh}\!\left(\frac{Z'}{t'}\right)
 \right]_{\partial Q}
 +I_{\mathrm{corner}}.
 \label{eq:unfixed-mini-first-derivative}
\end{align}
The nonsmooth intersection of the EOW brane with the regulated asymptotic boundary requires the Hayward corner term~\cite{Hayward:1993my}. The outward normal vector to the brane is
\be
n_\mu=\frac{1}{Z\sqrt{t'^2-Z'^2}}(Z',-t',0).
\ee
The outward normal vector to the asymptotic boundary is
\be
m_\mu=\left(0,-\frac{1}{Z},0\right).
\ee
The inner product of these vectors is given by
\be
m\cdot n=\frac{t'}{\sqrt{t'^2-Z'^2}}
\ee
Since the rapidity $\alpha$ is defined by $\tanh \alpha=\frac{Z'}{t'}$, the inner product is rewritten as
\be
m\cdot n=\cosh \alpha.
\ee
This precisely cancels the endpoint term generated by the integration
by parts in \eqref{eq:unfixed-mini-first-derivative} since the corner term gives
\be
I_{\rm corner}=\frac{1}{8\pi G}\int dx \sqrt{q} \alpha =\frac{L_x}{8\pi G}\frac{1}{Z}\text{arctanh} \left(\frac{Z'}{t'}\right),
\ee
where $q$ is the determinant of the metric at the  corner.
The resulting
first-order action is therefore
\begin{align}
 I_{\mathrm{mini}}
 =
 \frac{L_x}{8\pi G_N}
 \int_{\sigma_i}^{\sigma_f}d\sigma\,
 \left[
 \frac{t'}{Z^2}
 -\frac{Z'}{Z^2}
 \operatorname{arctanh}
 \left(\frac{Z'}{t'}\right)
 +\frac{\phi'^{\,2}}
 {\sqrt{t'^{\,2}-Z'^{\,2}}}
 -\frac{\sqrt{t'^{\,2}-Z'^{\,2}}}{Z^2}V(\phi)
 \right],
\end{align}
up to contributions from fixed regulator surfaces that are independent
of the brane trajectory.  This cancellation removes only the rapidity
endpoint term associated with the second-derivative form of the
extrinsic-curvature action.  The ordinary canonical endpoint
variations remain and must be treated by specifying the boundary data.

Let $\mathcal L_{\mathrm{mini}}$ denote the integrand in the first line of
\eqref{eq:unfixed-mini-first-derivative}.  Variation with respect to the scalar
gives
\begin{equation}
 2\frac{d}{d\sigma}
 \left(
 \frac{\phi'}{\sqrt{t'^{\,2}-Z'^{\,2}}}
 \right)
 +\frac{\sqrt{t'^{\,2}-Z'^{\,2}}}{Z^2}V'(\phi)=0.
 \label{eq:unfixed-scalar-eom}
\end{equation}
Since $\mathcal L_{\mathrm{mini}}$ contains $t$ only through $t'$, the
embedding-time equation is a conservation law,
\begin{equation}
 \frac{d p_t}{d\sigma}=0,
 \quad
 p_t\equiv\frac{\partial\mathcal L_{\mathrm{mini}}}{\partial t'}.
 \label{eq:unfixed-pt-conservation}
\end{equation}
A direct differentiation yields
\begin{equation}
 p_t=
 \frac{t'^{\,2}}{Z^2\bigl(t'^{\,2}-Z'^{\,2}\bigr)}
 -\frac{t'}{Z^2\sqrt{t'^{\,2}-Z'^{\,2}}}V(\phi)
 -\frac{t'\phi'^{\,2}}
 {\bigl(t'^{\,2}-Z'^{\,2}\bigr)^{3/2}}.
 \label{eq:unfixed-pt}
\end{equation}

\subsection{Static gauge and the missing constraint}
\label{sec:static-gauge-missing-constraint}

We now impose the static gauge only after obtaining the gauge-unfixed field
equations,
\begin{equation}
 t=\sigma,\quad t'=1,\quad Z'=\dot Z,\quad\phi'=\dot\phi.
 \label{eq:static-gauge-choice}
\end{equation}
This is the condition that the brane profile depends on the function of time $Z=Z(t)$.
After the corner term has canceled the endpoint term in
\eqref{eq:unfixed-mini-first-derivative}, the gage-fixed action is
\begin{align}
 I_{\mathrm{mini}}
 =\frac{L_x}{8\pi G_N}\int dt\,
 \left[
 \frac{1}{Z^2}
 -\frac{\dot Z}{Z^2}\operatorname{arctanh}(\dot Z)
 +\frac{\dot\phi^2}{\sqrt{1-\dot Z^2}}
 -\frac{\sqrt{1-\dot Z^2}}{Z^2}V(\phi)
 \right].
 \label{eq:static-gauge-action}
\end{align}
The scalar equation becomes
\begin{equation}
 2\frac{d}{dt}
 \left(
 \frac{\dot\phi}{\sqrt{1-\dot Z^2}}
 \right)
 +\frac{\sqrt{1-\dot Z^2}}{Z^2}V'(\phi)=0.
 \label{eq:static-scalar-eom}
\end{equation}
The conserved momentum conjugate to the embedding time reduces to
\begin{equation}
 p_t=
 \frac{1}{Z^2(1-\dot Z^2)}
 -\frac{V(\phi)}{Z^2\sqrt{1-\dot Z^2}}
 -\frac{\dot\phi^2}{(1-\dot Z^2)^{3/2}}.
 \label{eq:static-pt}
\end{equation}
Let $\mathcal L_t$ denote the integrand of
Eq.~\eqref{eq:static-gauge-action}, with the overall factor
$L_x/(8\pi G_N)$ suppressed.  The parametrized and gauge-fixed descriptions
are related by
\begin{equation}
 p_t=-H_t,
 \quad
 H_t=\dot Z\frac{\partial\mathcal L_t}{\partial\dot Z}
 +\dot\phi\frac{\partial\mathcal L_t}{\partial\dot\phi}
 -\mathcal L_t.
 \label{eq:pt-minus-static-H}
\end{equation}
If the static gauge is imposed before variation, the $t$ equation is absent.
The remaining gauge-fixed equations imply only that the corresponding
Hamiltonian is conserved; they do not determine its value.  Writing
$p_t=C$, the equation retained from the gauge-unfixed theory is
\begin{equation}
 \sqrt{1-\dot Z^2}
 -(1-\dot Z^2)V(\phi)
 -Z^2\dot\phi^2
 =C Z^2(1-\dot Z^2)^{3/2}.
 \label{eq:static-general-energy-constant}
\end{equation}
The original EOM for the brane requires 
$C=0$.  Equivalently,
\begin{equation}
 V(\phi)
 =\frac{1}{\sqrt{1-\dot Z^2}}
 -\frac{Z^2\dot\phi^2}{1-\dot Z^2},
 \label{eq:missing-constraint-explicit}
\end{equation}
or
\begin{equation}
 \frac{Z^2\dot\phi^2}{\sqrt{1-\dot Z^2}}
 +\sqrt{1-\dot Z^2}\,V(\phi)=1.
 \label{eq:missing-constraint-energy-form}
\end{equation}
Combining \eqref{phid2_Z} and \eqref{V_Z} with $\epsilon=-1$
reproduces \eqref{eq:missing-constraint-energy-form}.  In the gauge-fixed
minisuperspace theory, this relation is the missing Hamiltonian constraint.

\subsubsection*{On-shell reduction and its limitations}
\label{sec:onshell-reduction-limitations}

For later use in the semiclassical wavefunction, it is helpful to state how
the boundary on-shell action emerges from the gauge-unfixed formulation.  In
static gauge, before integrating the extrinsic-curvature term by parts, the
action takes the form
\begin{align}
 I_{\mathrm{mini}}
 =\frac{L_x}{8\pi G_N}\int dt\,
 \left[
 \frac{1}{Z^2}
 -\frac{\ddot Z}{Z(1-\dot Z^2)}
 +\frac{\dot\phi^2}{\sqrt{1-\dot Z^2}}
 -\frac{\sqrt{1-\dot Z^2}}{Z^2}V(\phi)
 \right]
 +I_{\mathrm{corner}}.
 \label{eq:static-second-derivative-action}
\end{align}
Notice that here we keep the second-derivative term. For the convention adopted here,
\begin{equation}
 I_{\mathrm{corner}}
 =\frac{L_x}{8\pi G_N}
 \left[
 \frac{1}{Z}\operatorname{arctanh}(\dot Z)
 \right]_{\partial Q}.
 \label{eq:poincare-corner-term}
\end{equation}
Using EOMs, one finds
\begin{equation}
 \frac{\dot\phi^2}{\sqrt{1-\dot Z^2}}
 -\frac{\sqrt{1-\dot Z^2}}{Z^2}V(\phi)
 =-\frac{1}{Z^2}
 +\frac{\ddot Z}{Z(1-\dot Z^2)}.
 \label{eq:onshell-cancellation}
\end{equation}
All the bulk terms in \eqref{eq:static-second-derivative-action} therefore cancel,
and the classical action reduces to
\begin{equation}
 I_{\mathrm{cl}}
 =\frac{L_x}{8\pi G_N}
 \left[
 \frac{1}{Z}\operatorname{arctanh}(\dot Z)
 \right]_{\partial Q}.
 \label{eq:onshell-boundary-result}
\end{equation}
Notice that this reduction does not define an off-shell action for two-dimensional
gravity. 

\subsection{Exact gauge-fixed brane action from integrating out the AdS$_3$ bulk}
\label{subsec:exact-gf-brane-action}

In this subsection we derive the exact two-dimensional action obtained by
integrating out the AdS$_3$ bulk, before making any derivative expansion.
We work with AdS radius set to one and use the Poincare metric
\begin{equation}
ds^2
=
\frac{dZ^2-dT^2+dX^2}{Z^2}.
\end{equation}
The bulk cosmological constant is  $\Lambda=-1 $, so that on shell
\begin{equation}
R_{\mu\nu}=-2g_{\mu\nu},
\quad
R=-6,
\quad
R-2\Lambda=-4.
\end{equation}
The starting point is the regulated action
\begin{align}
I
&=
\frac{1}{16\pi  G}
\int_{M} d^3x\,\sqrt{-g}\,(R-2\Lambda)
+
\frac{1}{8\pi  G}
\int_{N} d^2\xi\,\sqrt{-h}\,K\nn\\
&+
\frac{1}{8\pi  G}
\int_{Q} d^2\xi\,\sqrt{-h}
\left(K-
h^{ab}\partial_a\phi\partial_b\phi+V(\phi)
\right)
+
I_{\Sigma}.
\end{align}
Here  $Q $ is the end-of-the-world brane,  $h_{ab} $ is the induced
metric on  $Q $,  $K $ is the trace of its extrinsic curvature, and
 $I_{\Sigma} $ denotes the
corner term. 

We now choose the general brane profile
\begin{equation}
T=t,
\quad
X=x,
\quad
Z=Z(t,x).
\end{equation}
This is a gauge fixing of the worldvolume diffeomorphisms. It is valid in
a patch where
\begin{equation}
\det\frac{\partial(T,X)}{\partial(\xi^0,\xi^1)}\neq0.
\end{equation}
Equivalently, the brane is represented  over the boundary
coordinates  $(T,X) $. 

The embedding is
\begin{equation}
X^\mu(t,x)
=
\bigl(T,Z,X\bigr)
=
\bigl(t,Z(t,x),x\bigr).
\end{equation}
The tangent vectors are
\begin{equation}
E^\mu_t
=
\partial_tX^\mu
=
(1,Z_t,0),
\quad
E^\mu_x
=
\partial_xX^\mu
=
(0,Z_x,1).
\label{tangent}
\end{equation}
The induced metric is therefore
\begin{equation}
h_{tt}
=
\frac{-1+Z_t^2}{Z^2},
\quad
h_{tx}
=
\frac{Z_tZ_x}{Z^2},
\quad
h_{xx}
=
\frac{1+Z_x^2}{Z^2}.
\end{equation}
Thus, we have
\begin{equation}
h_{ab}
=
\frac{1}{Z^2}
\begin{pmatrix}
-1+Z_t^2 & Z_tZ_x \\
Z_tZ_x & 1+Z_x^2
\end{pmatrix},
\end{equation}
and
\begin{equation}
\sqrt{-h}
=
\frac{\sqrt{1-Z_t^2+Z_x^2}}{Z^2}.
\end{equation}
The inverse induced metric is
\begin{equation}
h^{ab}
=
\frac{Z^2}{1-Z_t^2+Z_x^2}
\begin{pmatrix}
-(1+Z_x^2) & Z_tZ_x \\
Z_tZ_x & 1-Z_t^2
\end{pmatrix}.
\end{equation}
We choose the bulk region to be  $z\geq Z(t,x) $ and the on-shell bulk term is simply a volume term:
\begin{equation}
I_{\rm bulk}
=
\frac{1}{16\pi  G}
\int_N d^3x\,\sqrt{-g}\,(R-2\Lambda)
=
-\frac{1}{4\pi  G}
\int_N d^3x\,\sqrt{-g}=-\frac{1}{8\pi G}
\int dt\,dx\,\frac{1}{Z^2}.
\end{equation}

Next, we compute the brane Gibbons--Hawking term. The brane is the level
set
\begin{equation}
F(T,Z,X)=Z-Z(T,X)=0.
\end{equation}
For the bulk region  $F\geq0 $, the outward-pointing unit normal vector is 
\begin{equation}
n_\mu
=
\frac{1}{Z\sqrt{1-Z_t^2+Z_x^2}}
\left(
Z_t,\,-1,\,Z_x
\right),
\end{equation}
where the components are ordered as  $(T,Z,X) $. 

The extrinsic curvature is computed directly as 
\be
K_{tt}
=
\frac{ZZ_{tt}-1+Z_t^2}
{Z^2\sqrt{1-Z_t^2+Z_x^2}},\quad K_{tx}
=
\frac{ZZ_{tx}+Z_tZ_x}
{Z^2\sqrt{1-Z_t^2+Z_x^2}},\quad K_{xx}
=
\frac{ZZ_{xx}+1+Z_x^2}
{Z^2\sqrt{1-Z_t^2+Z_x^2}}.
\ee
The derivation is summarized in appendix \ref{app:direct-K-components}. The trace of the extrinsic curvature is
\begin{equation}
K
=
\frac{
2(1-Z_t^2+Z_x^2)
-
Z\left[
(1+Z_x^2)Z_{tt}
+
(Z_t^2-1)Z_{xx}
-
2Z_tZ_xZ_{tx}
\right]
}
{
(1-Z_t^2+Z_x^2)^{3/2}
}.
\end{equation}
Combining the bulk and brane action, we obtain
\begin{align}
    I_{\rm bulk}+I_{\rm brane}=\frac{1}{8\pi G}\int dtdx&\left(\frac{
2(1-Z_t^2+Z_x^2)
-
Z\left[
(1+Z_x^2)Z_{tt}
+
(Z_t^2-1)Z_{xx}
-
2Z_tZ_xZ_{tx}
\right]
}
{
Z^2(1-Z_t^2+Z_x^2)
}
-
\frac{1}{Z^2}\right.\nn\\
&\left.+
\frac{
(1+Z_x^2)\phi_t^2
-
2Z_tZ_x\phi_t\phi_x
-
(1-Z_t^2)\phi_x^2
}
{
\sqrt{1-Z_t^2+Z_x^2}
}
-
\frac{\sqrt{1-Z_t^2+Z_x^2}}{Z^2}V(\phi)\right).\label{action}
\end{align}

It is useful to compare this result with the finite-cutoff effective action
derived in Ref.~\cite{Callebaut:2025thw}. In the absence of the brane scalar, $\phi=0$,
and for $V=1$, \eqref{action} agrees, up to total derivative terms, with
Eq.~(4.7) of Ref.~\cite{Callebaut:2025thw} upon identifying $ Z=e^{-(\bar\rho+\tilde\phi)}$.

\subsection{Slow-gradient expansion and Liouville-like variables}
\label{subsec:slow-liouville}

We now derive the local Liouville-like action by expanding the exact
 action at small brane velocity and small brane curvature.
Introduce
\begin{equation}
Z=e^{-\Phi}.
\end{equation}
We use
\begin{equation}
\eta_{ab}=\mathrm{diag}(-1,+1),
\quad
(\partial f)^2
\equiv
\eta^{ab}\partial_a f\partial_b f
=
-f_t^2+f_x^2,
\end{equation}
and
\begin{equation}
\Box
\equiv
\eta^{ab}\partial_a\partial_b
=
-\partial_t^2+\partial_x^2.
\end{equation}
The potential  $V(\phi) $ itself is not assumed to be small.

First, we can rewrite $Z$ in terms of $\Phi$
\begin{equation}
Z_t=-Z\Phi_t,
\quad
Z_x=-Z\Phi_x,
\end{equation}
and 
\begin{equation}
1-Z_t^2+Z_x^2
=
1+Z^2(\partial\Phi)^2
=
1+e^{-2\Phi}(\partial\Phi)^2.
\end{equation}
Hence
\begin{equation}
\sqrt{1-Z_t^2+Z_x^2}
=
1+\frac12 e^{-2\Phi}(\partial\Phi)^2
+
O(\partial^4),
\end{equation}
and
\begin{equation}
\frac{1}{1-Z_t^2+Z_x^2}
=
1-e^{-2\Phi}(\partial\Phi)^2
+
O(\partial^4).
\end{equation}
We can expand (\ref{action}) up to the second order and we have
\begin{equation}
I_L=I_{\rm bulk}+I_{\rm brane}
=
\frac{1}{8\pi  G}
\int dt\,dx
\left[
e^{2\Phi}\left(1-V(\phi)\right)
+
\left(1-\frac12V(\phi)\right)
\eta^{ab}\partial_a\Phi\partial_b\Phi
-
\eta^{ab}\partial_a\phi\partial_b\phi
\right]
+
O(\partial^4).
\end{equation}
This is the Liouville-like action coupled to the scalar
 $\phi $. 

Varying  $I_L $ with respect to  $\Phi $ gives
\begin{equation}
E_\Phi\equiv\partial_a
\left[
\left(1-\frac12V(\phi)\right)\partial^a\Phi
\right]
-
e^{2\Phi}\left(1-V(\phi)\right).\label{EOMPHI}
\end{equation}
Varying with respect to  $\phi $, we find
\begin{equation}
E_\phi\equiv2\Box\phi
-
e^{2\Phi}V'(\phi)
-
\frac12V'(\phi)(\partial\Phi)^2
=0.\label{EOMscalar}
\end{equation}

If  $\phi=\phi_0 $ is a constant scalar saddle with
\begin{equation}
V'(\phi_0)=0,
\end{equation}
then the  $\Phi $ equation reduces to the Liouville equation
\begin{equation}
\Box\Phi
=
\frac{1-V(\phi_0)}{1-\frac12V(\phi_0)}
e^{2\Phi}.
\end{equation}

To clarify the potential dependence of the EOM, we left the potential explicitly below, but in our approximation $|\partial Z|\ll1$, the brane should be placed at the asymptotic boundary or the potential should be $V\sim1$. Notice that in this case, the Liouville field $\Phi$  has a wrong kinetic sign and the theory reduces to the timelike Liouville theory. Refer to \cite{Harlow:2011ny} for the timelike Liouville theory.

\subsection{Relation between the Neumann condition and the Liouville and scalar equations}
\subsubsection{Exact equations of motion and constraints}
\label{subsec:exact-eom-constraints}

Before delving into the derivative expansion, we clarify the relation between the EOM for the induced metric and scalar field obtained by varying the original brane action. They are not independent.

The original brane EOM is 
\begin{equation}
\mathcal N_{ab}=0,
\end{equation}
where
\begin{equation}
\mathcal N_{ab}
\equiv
K_{ab}
-
K h_{ab}
+
h_{ab}
\left(
h^{cd}\partial_c\phi\partial_d\phi+V(\phi)
\right)
-
2\partial_a\phi\partial_b\phi.
\end{equation}
This is a symmetric tensor equation on the two-dimensional brane and
therefore contains three components.
The scalar EOM is
\begin{equation}
\mathcal E_\phi
\equiv
2\nabla_h^2\phi
-
V'(\phi)
=
0,
\end{equation}
where  $\nabla_h^2=h^{ab}\nabla_a\nabla_b $ is the scalar Laplacian associated with
 $h_{ab} $. In density form this equation is
\begin{equation}
2\partial_a
\left(
\sqrt{-h}\,h^{ab}\partial_b\phi
\right)
-
\sqrt{-h}\,V'(\phi)
=
0.
\end{equation}

The brane EOM  and the scalar equation are related by
\begin{equation}
\nabla^a\mathcal N_{ab}
+
\mathcal E_\phi\,\partial_b\phi
=
0.
\label{codda}
\end{equation}
Indeed, the Codazzi equation gives
\begin{equation}
\nabla^a(K_{ab}-Kh_{ab})
=
R_{\mu\nu}n^\mu E^\nu_b
=
0,
\end{equation}
because $R_{\mu\nu}\propto g_{\mu\nu}$ and  $n_\mu E^\mu_b=0 $. $ E$ is the tangent vector defined in \eqref{tangent}.
This is a conservation of stress-energy tensor. For the scalar
part,
\begin{equation}
\nabla^a
\left[
h_{ab}\left((\nabla\phi)^2+V(\phi)\right)
-
2\nabla_a\phi \nabla_b\phi
\right]
=
\left(
V'(\phi)-2\nabla^2\phi
\right)\nabla_b\phi=
-\mathcal E_\phi\,\nabla_b\phi.
\end{equation}
Thus, we prove the above relation \eqref{codda}. 

\subsubsection{Consistency of equations of motion}
Next, we clarify the relation between the Liouville-like equation, the
scalar equation, and the original Neumann condition. The purpose of this
subsection is to show that the slow-gradient equations obtained from
$I_L$, supplemented by the two constraint equations, reproduce the
$O(\partial^2)$ expansion of the original brane equations.

The scalar EOM of the brane is 
\begin{equation}
\mathcal{E}_\phi=2\partial_a
\left(
\sqrt{-h}\,h^{ab}\partial_b\phi
\right)
-
\sqrt{-h}\,V'(\phi)
=0.
\end{equation}
In the slow-gradient expansion, we find
\begin{equation}
\sqrt{-h}\,h^{ab}\partial_b\phi
=
\eta^{ab}\partial_b\phi
+
O(\partial^3),
\end{equation}
which leads to
\begin{equation}
0=2\partial_a
\left(
\sqrt{-h}\,h^{ab}\partial_b\phi
\right)
-
\sqrt{-h}\,V'(\phi)
=
2\Box\phi
-
e^{2\Phi}V'(\phi)
-
\frac12V'(\phi)(\partial\Phi)^2
+
O(\partial^4).
\end{equation}
Thus, (\ref{EOMscalar}) is reproduced up to the  second-order expansion of the
 scalar equation.

We next expand the EOM for the extrinsic curvature of the brane
\begin{equation}
\mathcal{N}_{ab}
=
K_{ab}
-
K h_{ab}
+
h_{ab}
\left(
h^{cd}\partial_c\phi\partial_d\phi+V(\phi)
\right)
-
2\partial_a\phi\partial_b\phi .
\end{equation}
To second order in derivatives, the induced metric is
\begin{equation}
h_{ab}
=
e^{2\Phi}\eta_{ab}
+
\partial_a\Phi\partial_b\Phi
+
O(\partial^4),
\end{equation}
and its inverse is
\begin{equation}
h^{ab}
=
e^{-2\Phi}\eta^{ab}
-
e^{-4\Phi}\partial^a\Phi\partial^b\Phi
+
O(\partial^4).
\end{equation}
The extrinsic curvature is expressed as
\begin{equation}
K_{ab}
=
e^{2\Phi}\eta_{ab}
-
\partial_a\partial_b\Phi
+
2\partial_a\Phi\partial_b\Phi
-
\frac12\eta_{ab}(\partial\Phi)^2
+
O(\partial^4),
\end{equation}
and its trace is similarly given by
\begin{equation}
K
=
2
-
e^{-2\Phi}\Box\Phi
+
O(\partial^4).
\end{equation}
Substituting these expressions into $\mathcal{N}_{ab}$, we find
\begin{equation}
\begin{aligned}
\mathcal{N}_{ab}
=
&
-\partial_a\partial_b\Phi
+
\eta_{ab}\Box\Phi
+
e^{2\Phi}\left(V(\phi)-1\right)\eta_{ab}
\\
&
-\frac12\eta_{ab}(\partial\Phi)^2
+
\eta_{ab}(\partial\phi)^2
+
V(\phi)\partial_a\Phi\partial_b\Phi
-
2\partial_a\phi\partial_b\phi
+
O(\partial^4).
\end{aligned}
\label{eq:2ndNeumann}
\end{equation}
We can explicitly write down the each component as
\ba
\mathcal{N}_{tt}&=& -(\dot{\phi}^2+\phi'^2)+\frac{1}{2}\left(\dot{\Phi}^2+\Phi'^2\right)-\Phi''-e^{2\Phi}(V-1),\no
\mathcal{N}_{xx}&=& -(\dot{\phi}^2+\phi'^2)+\frac{1}{2}\left(\dot{\Phi}^2+\Phi'^2\right)-\ddot{\Phi}+e^{2\Phi}(V-1),\no
\mathcal{N}_{tx}&=& -2\dot{\phi}\phi'^2+\dot{\Phi}\Phi'-\dot{\Phi}',
\ea
where $\dot{f}=\frac{\partial f}{\partial t}$ and $f'=\frac{\de f}{\de x}$. Below, in our approximation, we utilize $V\sim 1$ at the leading order.

First, let us comment on the relation between the EOM for $\Phi$ and the original brane EOM. Raising
both indices of $K_{ab}$ with $h^{ab}$, we obtain
\begin{equation}
K^{ab}
=
e^{-2\Phi}\eta^{ab}
-
e^{-4\Phi}\partial^a\partial^b\Phi
-
\frac12e^{-4\Phi}\eta^{ab}(\partial\Phi)^2
+
O(\partial^4).
\end{equation}
Contracting this with $\mathcal{N}_{ab}$ gives
\begin{equation}
\frac12 e^{2\Phi}K^{ab}\mathcal{N}_{ab}
=
\left(1-\frac12V(\phi)\right)\Box\Phi
-
e^{2\Phi}\left(1-V(\phi)\right)
+
O(\partial^4).
\end{equation}
Comparing this expression with (\ref{EOMPHI}), we find
\begin{equation}
E_\Phi
=
\frac12 e^{2\Phi}K^{ab}\mathcal{N}_{ab}
+
\frac12e^{-2\Phi}
\left(
\partial_a\Phi\,\partial^a\phi
\right)
E_\phi
+
O(\partial^4).
\end{equation}
Therefore, once the scalar equation $E_\phi=0$ is imposed, the
Liouville-like equation $E_\Phi=0$ is equivalent to
\begin{equation}
K^{ab}\mathcal{N}_{ab}=0
\end{equation}
up to $O(\partial^4)$.

In our gauge, the two worldvolume diffeomorphisms have already
been fixed. Hence, the two components of the Neumann condition are \textit{not}
obtained by varying the gauge-fixed local action. These should be imposed by hand as constraints additionally. Let us turn to $\mathcal{N}_{ab}$.
Indeed we find that $\mathcal{N}_{tt}-\mathcal{N}_{xx}=\ddot{\Phi}-\Phi''-2e^{\Phi}(V-1)$ vanishes due to the Liouville equation of motion. After using the EOM for $\Phi$, we can also confirm
\ba
\de^a \mathcal{N}_{ab}=(\de_b\phi)\cdot \left(-2\de^2\phi+V'e^{2\Phi}\right),
\ea
is vanishing owing to the equation of motion of $\phi$.

When we assume the translationally invariant setup i.e. $\phi$ and $\Phi$ are independent of $x$, we have $\mathcal{N}_{tx}=0$ and the conservation law  $\de^a \mathcal{N}_{ab}=0$ tells us that $\mathcal{N}_{tt}=-\dot{\phi}^2+\frac{1}{2}\dot{\Phi}^2-e^{2\Phi}(V-1)\simeq \mathcal{N}_{xx}$ does not depend also on $t$. 
Thus, the extra condition we have to impose, other than the equations of motion for $\phi$ and $\Phi$, is only this Hamiltonian constraint $\mathcal{N}_{tt}=0$ at an arbitrarily chosen time. 

Now we turn to the generic case where $\phi$ and $\Phi$ depends on both $t$ and $x$. The field equations of motion for $\phi$ and $\Phi$ lead to $\mathcal{N}_{tt}-\mathcal{N}_{xx}=0$ and $\de^a \mathcal{N}_{ab}=0$ as we mentioned. This leads to $(\de^2_t-\de^2_x)\mathcal{N}_{ab}=0$ and  is solved as
\ba
&& \mathcal{N}_{tt}=\mathcal{N}_{xx}=A(t-x)+B(t+x),\no
&& \mathcal{N}_{tx}=-A(t-x)+B(t+x),
\ea
where $A$ and $B$ are arbitrary functions. To correctly reproduce the brane equation of motion, we need to impose $A=B=0$ or equally setting $\mathcal{N}_{tt}=\mathcal{N}_{tx}=0$ at an arbitrary chosen time by hand in the Liouville field theory.

In conclusion, the complete second-order
Liouville-like system is described by
\begin{equation}
E_\Phi=0,
\quad
E_\phi^{L}=0,
\quad
\mathcal{N}_{tt}|_{t=t_0}=0,
\quad
\mathcal{N}_{tx}|_{t=t_0}=0,
\end{equation}
where $t_0$ is a specific time which we can choose freely. This system is equivalent, to the same order in the derivative expansion,
to the brane system with equations of motion
\begin{equation}
\mathcal{N}_{ab}=0,
\quad
E_\phi=0.
\end{equation}

For the derivation of the full constraints, we must keep all the diffeomorphisms or take a variation of the induced metric, see \cite{Neuenfeld:2024gta} for a  similar  spirit.


\subsection{Covariant description in terms of two-dimensional gravity}
As the final analysis of the basic dynamics of the brane, let us consider a covariant description as a two-dimensional gravity. To see this, let us note that $\mathcal{N}_{ab}$ is decomposed as
\ba
\mathcal{N}_{ab}=T^{(\phi)}_{ab}+T^{(\Phi)}_{ab},
\ea
where we introduced 
\ba
&& T^{(\phi)}_{ab}=\eta_{ab}(\de\phi)^2-2\de_a\phi\de_b\phi,\no
&& T^{(\Phi)}_{ab}=\de_a\Phi\de_b\Phi-\frac{1}{2}\eta_{ab}(\de\Phi)^2-\de_a\de_b\Phi+\eta_{ab}(\de^2\Phi)+e^{2\Phi}(V-1)\eta_{ab}.
\ea
This is consistent with \eqref{eq:2ndNeumann} since $V\sim 1$.
Notice that $T^{(\phi)}_{ab}$ coincides with the energy stress tensor of a massless free boson, while  $T^{(\Phi)}_{ab}$ does with that of a Liouville field with the potential $e^{2\Phi}(V-1)$. A useful fact is that the covariant two-dimensional gravity  is equivalent to the Liouville field theory, is given by the Polyakov action \cite{Polyakov:1981rd}
by introducing the two-dimensional metric by
\ba
h_{ab}=\eta_{ab}e^{2\Phi}.  \label{anz}
\ea
In summary we find that the brane model in the small slope approximation is equivalent to the following two-dimensional gravity
\ba
I_{\mbox{2d gravity}}=\frac{1}{8\pi G_N}\int dtdx \s{-h}\left[-\frac{1}{8}R\frac{1}{\Box}R-h^{ab}\de_a\phi\de_b\phi-(V(\phi)-1)\right].
\ea
By taking the variation of $h_{ab}$, we clearly obtain $\mathcal{N}_{ab}=T^{(\phi)}_{ab}+T^{(\Phi)}_{ab}=0$ assuming the conformal gauge (\ref{anz}). Though in addition we obtain the equation of motion for the scalar field $\phi$, it is not independent from $\mathcal{N}_{ab}=0$ as we have seen. In this way we can rewrite the brane dynamics after integrating out the AdS$_3$ in terms of two-dimensional gravity. Though this was already expected in \cite{Suzuki:2022xwv,Fujiki:2025yyf}, here we presented a direct and complete derivation.

\section {Brane inflation in Poincar\'e coordinates}\label{sec:Inflation}

\subsection{Review of the slow-roll inflation}
Cosmological observations show that our universe did not start from the hot Big Bang. Before that, there was an accelerated expansion period called inflation, which can be approximated by de Sitter space. At an early time, de Sitter is a good approximation for inflation. However, at late times, de Sitter solution must be violated in order to have a transition to the hot Big Bang.

Consider an expanding universe metric $ds^2 = -dt^2 + a(t)^2 dx^2 $, where $a(t)$ is the scale factor. De Sitter space is characterized by a constant expansion rate $H \equiv \dot{a} / a = \mathrm{const}$. A small deviation from the de Sitter solution can be parametrized by the slow-roll (SR) parameter, which will be defined precisely later.
A textbook realization of inflation is a single scalar field $\phi$ slow-rolling down a potential $V(\phi)$ in Einstein gravity. This model is called canonical slow-roll inflation. The action of the inflation is
\begin{equation}
S = \frac{1}{2} \int \mathrm{d}^4 x \sqrt{-g} \left[ \mpl^2 R - \partial_\mu \phi \partial^\mu \phi - 2V(\phi) \right], \label{actionb}
\end{equation}
where $g = \mathrm{det} ~g_{\mu \nu}$, $g_{\mu \nu}$ is the metric, and $R$ is its Ricci scalar. The stress-energy tensor of the scalar field is 
\begin{equation}
T_{\mu \nu} = \frac{-2}{\sqrt{-g}} \frac{\delta S_\phi}{\delta g^{\mu \nu}} = \partial_\mu \phi \partial_\nu \phi - g_{\mu \nu} \left[ \frac{1}{2} \partial_\rho \phi \partial^\rho \phi + V(\phi) \right],
\end{equation}
and the energy density and pressure are
\begin{equation}
\rho_\phi = \frac{1}{2} \dot{\phi}^2 + V(\phi) ~~\mathrm{and}~~ p_\phi = \frac{1}{2} \dot{\phi}^2 - V(\phi),
\end{equation}
so the equation of state is
\begin{equation}
w_\phi = \frac{p_\phi}{\rho_\phi} = \frac{\dot{\phi}^2 - 2V(\phi)}{\dot{\phi}^2 + 2V(\phi)},
\end{equation}
which shows that requirement $w < -1/3$ is possible if the potential energy dominates over the kinetic energy.
The equation of motion of $\phi$ is the Klein-Gordon equation
\begin{equation}
\ddot{\phi} + 3 H \dot{\phi} + \frac{d V}{d\phi} = 0. \label{klein}
\end{equation}
and the equation of motion of $a$ is the Friedmann equation
\begin{equation}
H^2 = \frac{1}{3\mpl^2}\left(\frac{1}{2} \dot{\phi}^2 + V(\phi)\right), ~\dot{H} = - \frac{\dot{\phi}^2}{2 \mpl^2}.
\end{equation}

During inflation, the evolution of the Hubble parameter is parameterized by the SR parameters that is defined as
\begin{equation}
\epsilon_{n+1} = \frac{\dot{\epsilon}_n}{\epsilon_n H},  ~ \epsilon_1 \equiv - \frac{\dot{H}}{H^2} = \frac{\dot{\phi}^2}{2 \mpl^2 H^2}, \label{srpar}
\end{equation}
where $n$ is a positive integer. The equality for $\epsilon_1$ is obtained by substituting the Friedmann equation. In case the Hubble parameter $H$ is almost constant, we have $\epsilon_1 \ll 1$. This condition is called the quasi-de Sitter\index{De Sitter} approximation. Thus, at zeroth order in SR approximation, the scale factor can be written as
\begin{equation}
a(t) \propto e^{H t}.
\end{equation}
With the SR approximation, the equation of motion becomes
\begin{align}
H^2 & \simeq \frac{1}{3} V(\phi) \approx \mathrm{constant}, \label{eq:HtoV}\\
\dot{\phi} & \simeq - \frac{1}{3H} \frac{d V}{d\phi}. \label{eq:phidottoV'}
\end{align}

The conformal time for exactly constant $H$ is
\begin{equation}
\eta = \frac{1}{H} \int \frac{d a}{a^2} = - \frac{1}{aH} + \mathrm{const},
\end{equation}
where the boundary condition $a\rightarrow \infty$ is taken when $\eta \rightarrow 0$, so $\mathrm{const}=0$. In a realistic inflationary universe, the scale factor $a$ grows very large but not to infinity because the time independence of $H$ is merely an approximation. Including the first slow-roll correction, the scale factor is given by
\begin{equation}
a(\eta) \propto \frac{1}{(-\eta)^{1 + \epsilon_1}}, 
\end{equation}
thus the metric can be written as
\begin{equation}
ds^2 = \frac{1}{\eta^{2 + 2\epsilon_1}} (-d\eta^2 + dx^2).
\label{eq:inflationmetric}
\end{equation}
Inflation ends when the slow-roll condition is violated $\epsilon_1(\phi_{\mathrm{end}}) \simeq 1$. At the end of inflation, the scale factor $a_i$ is increased by a factor $e^{N_\mathrm{tot}}$ compared to the initial scale factor $a_s$, where $N_\mathrm{tot}$ is called the number of e-folds. $N_\mathrm{tot}$ can be expressed as
\begin{equation}
N_\mathrm{tot} = \log \frac{a_i}{a_s} = \int_{t_s}^{t_i} d t ~H = \int_{\phi_s}^{\phi_i} d \phi \frac{H}{\dot{\phi}} \approx \int_{\phi_i}^{\phi_s} d \phi \frac{V}{dV/d\phi},
\end{equation}
and to solve the horizon and flatness problem requires $N_\mathrm{tot} \geq 60$.

\subsection{Inflation solution in brane model}
\begin{figure}[h]
    \centering
    \includegraphics[ width=0.5\linewidth]{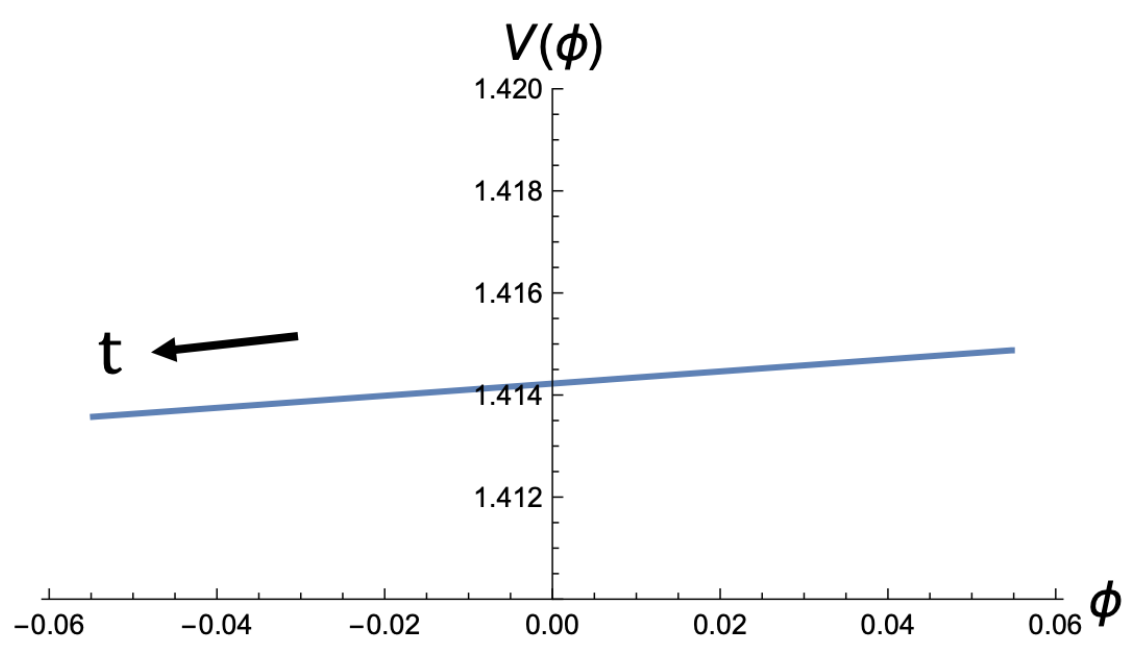}
    \caption{V-$\phi$ graph. We set the slow-roll parameter to be $\epsilon_1=10^{-4}$.}
    \label{fig:V-phi_inflation}
\end{figure}

We first consider the three-dimensional Poincar\'e coordinates:
\begin{equation}
    ds^2=\frac{dz^2-dt^2+dx^2}{z^2}. 
\end{equation}
To realize a slow-roll inflation on the EOW brane, we compare \eqref{eq:Poincareinduced} with \eqref{eq:inflationmetric}, and find the following conditions:
\begin{equation}
    Z(t)=\eta^{1+\ep_1},\quad \frac{d\eta}{\eta}=\frac{\sqrt{1-\left(\frac{dZ(t)}{dt}\right)^2}}{Z(t)}dt
\end{equation}
By solving them, we obtain the solution
\begin{equation}
    t=-Z(t)\, {}_2F_1\left[-\frac{1}{2},-\frac{1+\epsilon_1}{2\epsilon_1},1-\frac{1+\epsilon_1}{2\epsilon_1},-\frac{Z(t)^{-\frac{1+\epsilon_1}{2\epsilon_1}}}{(1+\epsilon_1)^2}\right].
\end{equation}
Here, $\epsilon_1$ is a slow-roll parameter, which induces the deformation from dS brane. The configuration is described as the hyperplane slightly bent inwards. This behavior guarantees that the null energy condition hold.
Expanding it to first order in $\epsilon_1$, we obtain
\begin{equation}
    t\simeq-\sqrt{2}\,Z+\frac{Z\log Z}{\sqrt{2}}\epsilon_1+\mathcal{O}(\epsilon_1^2).
\end{equation}
We invert this relation to express $Z(t)$:
\begin{equation}
    Z(t)\simeq-\frac{1}{\sqrt{2}}t-\frac{t}{2\sqrt{2}}\log \frac{(-t)}{\sqrt{2}}\epsilon_1 +\mathcal{O}(\epsilon_1^2).
\end{equation}
For this configuration, we can rewrite the Neumann boundary condition \eqref{eq:generalNBC} and solve them approximately as follows:
\begin{align}
    \phi(t)-\phi_*\simeq&\;\frac{\sqrt{\epsilon_1}}{2^{\frac{3}{4}}}\log (-t)+\frac{\epsilon_1^{\frac{3}{2}}}{4\cdot2^{\frac{3}{4}}}\log (-t)+O(\epsilon_1^{\frac{5}{2}}), \\
    V(t)\simeq&\;\sqrt{2}+\frac{\epsilon_1}{\sqrt{2}}\left(\log\frac{(-t)}{\sqrt{2}}+\frac{1}{2}\right) \no
    &\;+\frac{\epsilon_1^2}{4\sqrt{2}}\left(1-2\log 2+(\log 2)^2+4(1-\log 2)\log (-t)+4(\log (-t))^2\right)+O(\epsilon_1^3). 
\end{align}
We immediately find
\begin{equation}
    \begin{aligned}
    V(\phi) \simeq & \sqrt{2}+\frac{\epsilon_1}{2\sqrt{2}}(1-\log 2)+\frac{\epsilon_1^2}{4\sqrt{2}}(1-2\log 2+(\log 2)^2) \\
    &+2^{\frac{1}{4}}\sqrt{\epsilon_1}\left(1-\epsilon_1\left(\frac{3}{4}-\log 2\right)\right)(\phi-\phi_*)+2\epsilon_1(\phi-\phi_*)^2+O(\epsilon_1^{\frac{5}{2}}).
    \end{aligned}
\end{equation}
From \eqref{srpar}, \eqref{eq:HtoV} and \eqref{eq:phidottoV'}, we obtain the relation $(V'/V)^2\sim\epsilon_1$, which is consistent with this result.
We numerically solve the Neumann boundary condition and show $V-\phi$ graph in the Fig.~\ref{fig:V-phi_inflation}, which is also consistent with the usual slow-roll potential.

\section{On-shell action}
The goal of this section is to compute the semiclassical wavefunction
\begin{equation}
\Psi = e^{i I_\mathrm{cl}},
\end{equation}
where $I_\mathrm{cl}$ is the action evaluated at its classical solution with an appropriate boundary condition. We are interested in the probability
\begin{equation}
P = |\Psi|^2 = e^{-2 \mathrm{Im} I_\mathrm{cl}}.
\end{equation}
Therefore, we have to compute imaginary part of the on-shell action.

From the action in section~\ref{sec:BraneModel}
\begin{equation}
I = \frac{L_x}{8\pi G_N} \int dt \left[ \frac{1}{Z^2} - \frac{ \ddot{Z}}{Z (1 - \dot{Z}^2)} + \frac{\dot{\phi}^2}{\sqrt{1- \dot{Z}^2}} - \frac{1}{Z^2} \sqrt{1 - \dot{Z}^2} V(\phi) + \partial_t\left( \frac{1}{Z}  \tanh^{-1} \dot{Z} \right) \right]
\end{equation}
Substituting the classical equation of motion yields
\begin{equation}
I_\mathrm{cl} = \frac{L_x}{8\pi G_N} \left[ \frac{1}{Z}  \tanh^{-1} \dot{Z} \right]_\mathrm{bdy}
\end{equation}
For constant potential case $V = V_0$, the classical solutions are
\begin{align}
& \dot{Z}^2 + \frac{1}{(V_0 + p_\phi^2 Z^2)^2} = 1, \nonumber\\
& \dot{\phi} = \frac{p_\phi}{V_0 + p_\phi^2 Z^2},
\end{align}
where $p_\phi$ is an arbitrary constant. 

We would like to find a solution of the equation of motion that satisfies the Hartle-Hawking boundary condition. To do so, transform the equation of motion to Euclidean time $t \rightarrow - is$
\begin{equation}
\frac{1}{(V_0 + p_\phi^2 Z^2)^2} = 1 + (Z')^2,
\end{equation}
where prime denotes derivative with respect to $s$. First, consider case $p_\phi =0$, which corresponds to non-dynamical matter. The equation of motion becomes
\begin{equation}
1 + (Z')^2 = \frac{1}{V_0^2}.
\end{equation}
For $V_0 > 1$, there is no real solution for $Z'$. Then, consider case $p_\phi \neq 0$. At early time when $Z \gg 1$
\begin{equation}
1 + (Z')^2 \approx \frac{1}{(p_\phi^2 Z^2)^2} .
\end{equation}
Thus $1 + (Z')^2 \rightarrow 0$, which means we cannot have a real solution for $Z'$. We see that finding the Hartle-Hawking like solution in Poincare coordinate is difficult. Therefore, we will try to find it in the global coordinate, which cover the whole de Sitter spacetime.

\subsection{Action in global coordinate}
Here we study the equation of motion for the cosmological brane in a global AdS$_3$. Refer also to \cite{Fujiki:2025yyf} for an earlier work where the same system was analyzed and cosmological solutions are classified.
We fix the background metric to be
\begin{align}
    ds^2=d\rho^2-\cosh^2\rho\,d\tau^2+\sinh^2\rho\, d\varphi^2, 
    \label{gAdS}
\end{align}
where we have the periodicity $\vp\sim \vp+2\pi$.
We consider an EOW brane whose trajectory is given by $\rho=R(\tau)$, assuming rotational symmetry along the $\varphi$ direction. There are two possible choices for the bulk region:
\begin{align}
        \mathrm{Type \,I:} \,\,\,R(\tau) \leq \rho, \qquad
        \mathrm{Type \,II:} \,\,\,0\leq\rho \leq R(\tau). \label{domglobal}
\end{align}
We also assign $\epsilon=+1$ and $\epsilon=-1$ to Type I and Type II, respectively. The induced metric is given by
\begin{align}
    ds^2=-(\cosh^2 R-\dot{R}^2)d\tau^2+\sinh^2 R \,d\varphi^2, 
\end{align}
where $\dot{R}=\frac{dR(\tau)}{d\tau}$. We also impose the timelike condition $\cosh^2 R>\dot{R}^2$. In global coordinates, the extrinsic curvature is given by
\begin{equation}
K_{ab} dx^a dx^b = \epsilon \frac{\sinh R (2 \dot{R}^2 - \cosh^2 R) - \ddot{R} \cosh R }{\sqrt{\cosh^2 R - \dot{R}^2}} d\tau^2 - \epsilon \frac{\sinh R \cosh^2 R}{\sqrt{\cosh^2 R - \dot{R}^2}} d\varphi^2,
\end{equation}
thus the Neumann boundary condition can be rewritten as
\begin{align}
     \dot{\phi}^2&=\epsilon\frac{(1-\sinh^2R)\dot{R}^2+\cosh R(\ddot{R}\sinh R-\cosh R)}{2\sinh R\s{\cosh^2R-\dot{R}^2}},\\
     V(\phi)&=-\epsilon\frac{\cosh^2R(\cosh^2R+\sinh^2R)-\dot{R}^2(2\sinh^2R+\cosh^2R)+\ddot{R}\sinh R\cosh R }{2\sinh R(\cosh^2R-\dot{R}^2)^{3/2}}.  
\end{align}
The energy-conservation-like relation is given by
\begin{align}
    \f{\sinh{R}}{\cosh^2{R}}\left( \f{\dot{\phi}^2}{\sqrt{\cosh^2{R}-\dot{R}^2}}+V(\phi) \sqrt{\cosh^2{R}-\dot{R}^2} \right) =-\epsilon.
\end{align}

The action consists of the brane, bulk, and corner terms. The brane term is given by
\begin{equation}
I_\mathrm{brane} = \frac{1}{8\pi G_N}\int d\tau d\varphi \sqrt{-h} \left[ K -h^{\tau\tau} \dot{\phi}^2 - V(\phi) \right].
\end{equation}
The bulk term can be obtained by integrating the bulk action along $\rho$ direction within domain \eqref{domglobal} 
\begin{equation}
I_\mathrm{bulk} = \frac{1}{16\pi G_N}\int d\tau d\varphi d\rho \s{-g}(R-2\Lambda) = \frac{-1}{4\pi G_N}\int d\tau d\varphi d\rho \cosh\rho \sinh\rho = \frac{-1}{8\pi G_N}\int d\tau d\varphi (\cosh^2 R - 1)
\end{equation}
In principle, the corner term can be computed directly from the dot product of normal vectors. However, we will compute it indirectly by demanding the total action does not have fields with second derivative of time.

Substituting the explicit formula for extrinsic curvature, the total action reads
\begin{align}
I = \frac{1}{8\pi G_N}\int d\tau d\varphi &\left[ \frac{\ddot{R} \sinh R \cosh R  -\sinh^2 R (2 \dot{R}^2 - \cosh^2 R) }{\cosh^2 R - \dot{R}^2}  \right. \nonumber\\
&\left. + \frac{ \dot{\phi}^2 \sinh R}{\sqrt{\cosh^2 R - \dot{R}^2}} - \sinh R \sqrt{\cosh^2 R - \dot{R}^2} V(\phi) + 1 \right] + I_\mathrm{corner}.
\end{align}
We see that the bulk term contain second derivative of time $\ddot{R}$, which should be removed by the corner term. With the following algebraic properties
\begin{equation}
\partial_\tau\left( \sinh R \tanh^{-1}\frac{\dot{R}}{\cosh R} \right) = \frac{\ddot{R} \sinh R \cosh R - \dot{R}^2 \sinh^2 R}{\cosh^2 R - \dot{R}^2} + \dot{R} \cosh R \tanh^{-1}\frac{\dot{R}}{\cosh R} ,
\end{equation}
we can indirectly determine the corner term to be
\begin{equation}
I_\mathrm{corner} = \frac{-1}{8\pi G_N}\int d\tau d\varphi \partial_\tau \left( \sinh R \tanh^{-1}\frac{\dot{R}}{\cosh R} \right).
\end{equation}
Thus, the action becomes
\begin{align}
I = \frac{1}{8\pi G_N}\int d\tau d\varphi &\left[ \sinh^2 R - \dot{R} \cosh R \tanh^{-1}\frac{\dot{R}}{\cosh R} \right. \nonumber\\
&\left. + \frac{ \dot{\phi}^2 \sinh R}{\sqrt{\cosh^2 R - \dot{R}^2}} - \sinh R \sqrt{\cosh^2 R - \dot{R}^2} V(\phi) + 1 \right]
\end{align}

\subsection{Regularity requirement}
We are interested in a Hartle-Hawking-like solution, which is created from nothing in the Euclidean time. We write the Euclidean time as $s$ such that the universe is created at $s=0$ and the Euclidean time evolution ends at a time $s=s_0$. Then it is continued to the real time $\tau$ as $s=s_0+i\tau$. The induced metric of this Euclidean space looks like
\ba
ds^2=(\cosh^2 R+R'^2)ds^2 +\sinh^2 R d\varphi^2,
\ea
where we introduced $R'=\frac{dR(s)}{ds}$. Note the relation 
$R'^2=-\dot{R}^2$ at $s=s_0$. The equation of motion is now written in this Euclidean setup as follows:
\ba
&& \dot{\phi}^2=\frac{-(1-\sinh^2R)R'^2-\cosh R(R''\sinh R+\cosh R)}{2\sinh R\s{\cosh^2R+R'^2}},\label{gKEOMP}\\
&& V(\phi)=\frac{\cosh^2R(\cosh^2R+\sinh^2R)+R'^2(2\sinh^2R+\cosh^2R)-R''\sinh R\cosh R }{2\sinh R(\cosh^2R+R'^2)^{3/2}}.
\label{gEEOM}
\ea

We would like to impose the regularity at $s=0$ where the space is created from nothing, whose condition reads 
\begin{equation}
R(s)=A\s{s}+Bs^{\frac{3}{2}}+\ddd,   \label{asyma}
\end{equation}
in the limit $s\to 0$. We also set $\phi=0$ at $s=0$. Indeed, the induced metric approaches 
\ba
ds^2\simeq A^2\left(\frac{1}{4s}ds^2+sd\varphi^2\right), 
\ea
which is clearly smooth at $s=0$ by rewriting $s=\rho^2$, for any positive values of $A$. 

In order to have a smooth continuation from the Euclidean solution to the Lorentzian one at $s=s_0$ we need to impose
\begin{equation}
R'(s_0)=\phi'(s_0)=0.  \label{bouncec}
\end{equation}
By plugging this into (\ref{gEEOM}) we obtain the following condition at $s=s_0$:

\ba
R''(s_0)=-\ddot{R}(s_0)=-\frac{\cosh R_0}{\sinh R_0},\ \ \ V(s_0)=\frac{\cosh R_0}{\sinh R_0},
\ea
where we set $R(s_0)=R_0$.

\subsection{De Sitter solution}
Consider a constant potential $V(\phi) = 1/\tanh \xi$ that is parameterized by a constant $\xi$. Solution of the Neumann boundary condition is given by
\begin{equation}
\cosh R(\tau) = \frac{\cosh \xi}{\cos \tau} .
\end{equation}
Analytic continuation to Euclidean time $0 \leq s \leq \xi$ gives
\begin{equation}
\cosh R(s) = \frac{\cosh \xi}{\cosh (\xi - s)} , \label{dsbrsolp}
\end{equation}
which satisfies the regularity condition $R(0) = 0$. Taking the first derivative yields
\begin{equation}
R'(s) \sinh R(s) = \cosh \xi \frac{\sinh(\xi - s)}{\cosh^2(\xi - s)},
\end{equation}
which satisfies the smooth continuation condition $R'(s_0) = 0$ at $s_0 = \xi$.

Next, we would like to evaluate the on-shell action. Substituting Neumann boundary condition to the action yields
\begin{equation}
I = I_\mathrm{corner} + \frac{1}{4 G_N}\int d\tau,
\end{equation}
where we note that the angular integral $\int d\varphi =2\pi$. We will compute the on-shell action for two different cases below.

\subsubsection{Euclidean de Sitter}
We would like to compute probability of having a Euclidean de Sitter universe with size $R(s_*)$. First, we have to evaluate the corner term with the following boundary 
\begin{equation}
I_\mathrm{corner} = -\frac{1}{4 G_N} \left[ \sinh R(s) \tanh^{-1}\frac{iR'(s)}{\cosh R(s)} \right]_{s=0}^{s=s_*}.
\end{equation}
At the lower boundary, the corner term vanishes because
\begin{equation}
\sinh R(s) \rightarrow 0.
\end{equation}
Thus, the corner term is given by
\begin{equation}
I_\mathrm{corner} = -\frac{1}{4 G_N}  \sinh R(s_*) \tanh^{-1}\frac{i\sqrt{1 - \mathrm{sech}^2\xi \cosh^2 R(s_*) }}{\sinh R(s_*)} .
\end{equation}
Adding the time integral part, the on-shell action becomes
\begin{equation}
I = -\frac{1}{4G_N} \left[ i s_* + \sinh R(s_*) \tanh^{-1}\frac{i\sqrt{1 - \mathrm{sech}^2\xi \cosh^2 R(s_*) }}{\sinh R(s_*)} \right] .
\end{equation}

\subsubsection{Lorentzian de Sitter}
The universe undergoes nucleation from Euclidean time $s=0$ to $s = \xi$
\begin{equation}
I_\mathrm{corner} = -\frac{1}{4 G_N} \left[ \sinh R(s) \tanh^{-1}\frac{iR'(s)}{\cosh R(s)} \right]_{s=0}^{s=\xi}.
\end{equation}
Because $R'(s \rightarrow \xi) = 0$, the corner term vanishes. Therefore, the on-shell action reduces to
\begin{equation}
I = -\frac{1}{4 G_N} i \xi = -\frac{1}{4 G_N} i \tanh^{-1} \frac{1}{V}
\end{equation}

Thus the probability $P$ of nucleation of the dS brane looks like
\ba
\log P=\frac{1}{2 G_N}\tanh^{-1}\left[\frac{1}{V}\right].
\ea
If we note that the radius of the two-dimensional de Sitter space $R_{dS}$, which is identified with the brane world-volume, is related to $V$ via $V=\frac{\s{1+R_{dS}^2}}{R_{dS}}$, we obtain
\ba
\log P=\frac{1}{2 G_N}\tanh^{-1}\left[\frac{R_{dS}}{\s{1+R_{dS}^2}}\right].\label{qbraneP}
\ea
As is clear from (\ref{qbraneP}), the probability distribution prefers a dS with a large radius. This generalizes the celebrated argument \cite{RevModPhys.61.1} in four dimensions to two dimensions by assuming that the value of the potential $V$ is randomly chosen. 

If the two-dimensional gravity is treated as the classical gravity ignoring all quantum corrections, we expect to obtain the result in the large $R_{dS}$ limit:
\ba
\log P_0=\frac{1}{2 G_N}\log R_{dS}. \label{cbraneP}
\ea
This formula can be regarded as dS$_2$ version of de Sitter entropy $S_{dS_{d+1}}=\frac{R_{dS}^{d-1}}{4G_N}V_{d-1}$ for dS$_{d+1}$, which is identical to the logarithm of the probability. Here $V_{d-1}$ is the volume of a $d-1$ dimensional sphere with the unit radius. The derivation of dS entropy using de Sitter brane was also discussed in \cite{Hawking:2000da}. Thus we can regard the difference between (\ref{qbraneP}) and (\ref{cbraneP}) as the quantum corrections peculiar to our braneworld holography.

\subsection{Inflationary brane solution}
Now we consider the following brane profile by deforming (\ref{dsbrsolp}) by a parameter $\eta$:
\ba
\cosh R(s)=\left[\frac{\cosh\xi}{\cosh(\xi-\ap s)}\right]^\eta.
\ea
The Universe is created at $s=0$ and evolves until $s=s_0=\frac{\xi}{\ap}$ under the imaginary time. Then it continues to the Lorentzian time as $s=\frac{\xi}{\ap}+i\tau$.
For this solution, we find $\cosh R_0=(\cosh\xi)^\eta$. This solution satisfies the condition (\ref{asyma}) with
\ba
A=\s{2\ap\eta}\s{\tanh\xi}.
\ea
By imposing the smoothness condition (\ref{bouncec}), we obtain
\ba
\ap^2\eta=1.
\ea
Note that $\ap=\eta=1$ is the standard dS brane solution.

\begin{figure}[htbp]
		\centering
		\includegraphics[width=7cm]{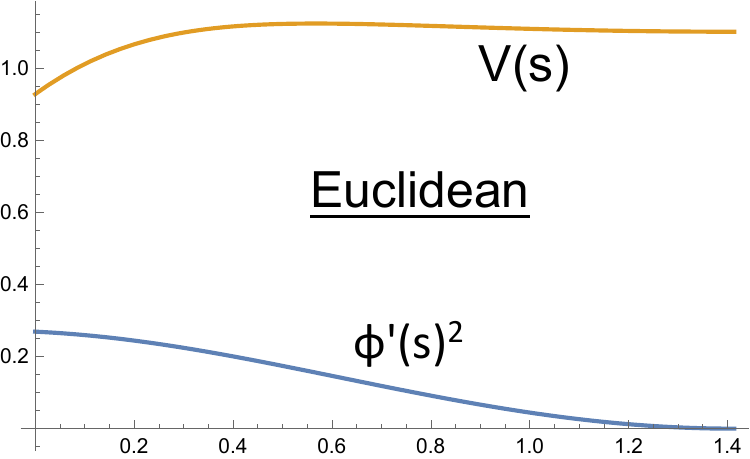}
        \includegraphics[width=7cm]{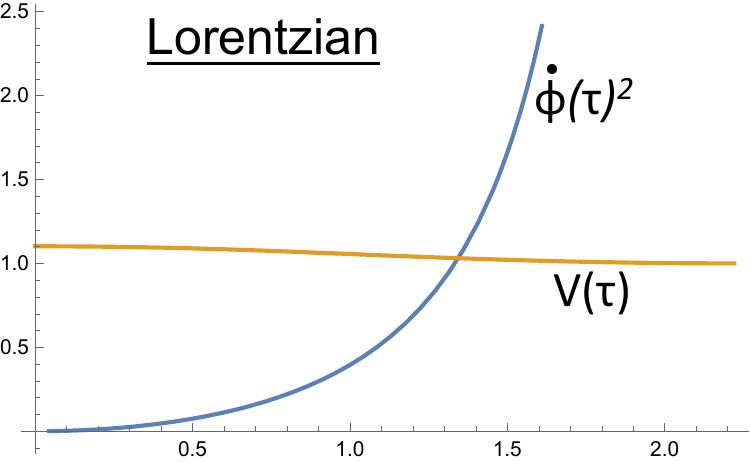}
		\caption{The plot of $\phi'^2$ (blue) and $V$ (orange) as a function of $s$ is shown in left panel. The plot of $\dot{\phi}^2$ (blue) and $V$ (orange) as a function of $\tau$ is shown in left panel. We set $\eta=2$, $\xi=1$ and $\ap=\frac{1}{\s{2}}$.} 
		\label{fig:Vpota}
\end{figure}

\begin{figure}[htbp]
		\centering
		\includegraphics[width=7cm]{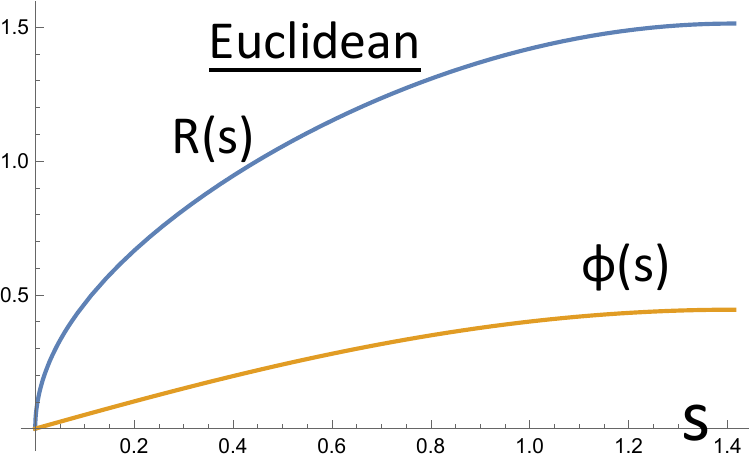}
        \includegraphics[width=7cm]{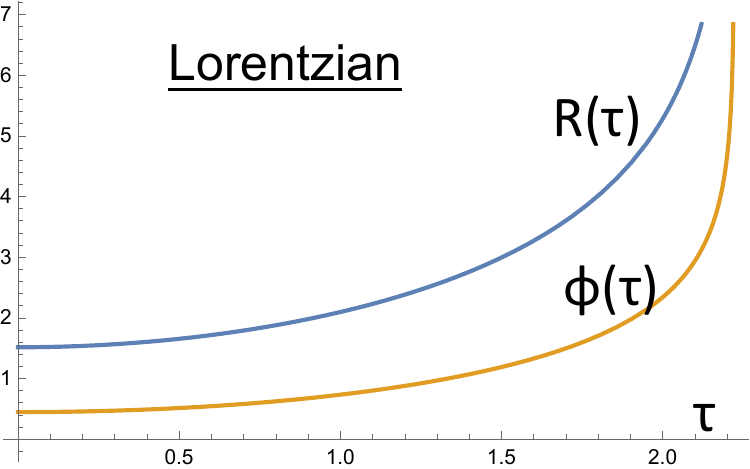}
		\caption{The plot of $R$ (blue) and $\phi$ (orange) as a function of $s$ (left panel) and $\tau$ (right panel). We set $\eta=2$, $\xi=1$ and $\ap=\frac{1}{\s{2}}$. We also set $\phi=0$ at $s=0$. We have $\phi(s_0)=0.44$.} 
		\label{fig:Vpotb}
\end{figure}

\begin{figure}[htbp]
		\centering
		\includegraphics[width=13cm]{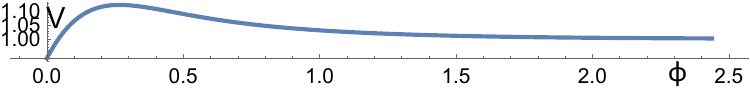}
		\caption{The plot of the potential $V(\phi)$ as a function of the scalar field $\phi$. We set $\eta=2$, $\xi=1$ and $\ap=\frac{1}{\s{2}}$. We also set $\phi=0$ at $s=0$. We have $\phi(s_0)=0.44$.} 
		\label{fig:Vpotc}
\end{figure}

In the special case $\xi=1$, $\eta=2$ and $\ap=\frac{1}{\s{2}}$, we numerically solve all relevant functions. They are plotted in Fig.\ref{fig:Vpota}, Fig.\ref{fig:Vpotb} and  Fig.\ref{fig:Vpotc}.

In the Lorentzian evolution we have
\ba
\cosh R(\tau)=\left[\frac{\cosh\xi}{\cos(\ap\tau)}\right]^\eta.
\ea
Note that $\tau$ takes values in the range 
$0\leq \tau<\frac{\pi}{2\ap}$. At $\tau=\frac{\pi}{2\ap}(\equiv\tau_0)$, we find $R=\infty$. It is also useful to study the late time behavior under the Lorentzian time evolution. By considering the limit 
$\tau\to \tau_0$, the induced metric on the brane behaves as
\ba
ds^2\simeq \frac{\cosh^2 R_0}{\left(\tau_0-\tau\right)^{2\eta}}(-d\tau^2+d\varphi^2),
\ea
which indeed looks like the inflation model by choosing $\eta=1+\ep_1$.

\section{Perturbation of brane}
In this section, we would like to check the stability of the brane under a small perturbation. We want to perturb the junction equation
\begin{equation}
K_{ab} - h_{ab} K = T_{ab} \equiv 2 \de_a\phi \de_b\phi - h_{ab} \left(h^{cd} \de_c\phi \de_d\phi + V(\phi) \right),
\end{equation}
and evaluate each component. Perturbation of the metric $Z(t,x)$ and scalar field $\phi(t,x)$ can be expressed as
\begin{align}
 Z(t,x) &= \bar{Z}(t) + f(t,x),\\
 \phi(t,x) &= \bar{\phi}(t) + \varphi(t,x) .
\end{align}

Next, we define
\begin{equation}
W \equiv h^{cd} \partial_c \phi \partial_d \phi + V(\phi) = W^{(0)} + W^{(1)},
\end{equation}
where $W^{(0)}$ and $W^{(1)}$ are $W$ evaluated on the background and its first-order perturbation, respectively. The explicit formula for the background term and its perturbation are
\begin{equation}
W^{(0)} = - \frac{1 - \dot{Z}^2 - Z\ddot{Z}}{(1 - \dot{Z}^2)^{3/2}},
\end{equation}
\begin{align}
W^{(1)} = -\frac{1}{Z (1- \dot{Z}^2)^{5/2}} &\left[ -2(1 - \dot{Z}^2)^2 f + Z\dot{Z}(1 - \dot{Z}^2) \partial_t f - Z^2(1 - \dot{Z}^2) \partial_t^2 f \right. \nonumber\\
&\left. -3 Z^2 \dot{Z} \ddot{Z} \partial_t f + Z \ddot{Z} (1 - \dot{Z}^2) f  \right].
\end{align}

Evaluating $tt$-component of the junction condition reads
\begin{equation}
K_{tt} -h_{tt}K = - \frac{\sqrt{1-\dot{Z}^2}}{Z^2} + \frac{1}{Z^3\sqrt{1-\dot{Z}^2}} \left[ 2(1 - \dot{Z}^2)f + Z\dot{Z} \partial_t f - Z^2(1 - \dot{Z}^2) \partial_x^2 f \right],
\end{equation}
\begin{equation}
T_{tt} = 2 \dot{\bar{\phi}}^2 + 4 \dot{\bar{\phi}} \partial_t \varphi + \frac{1 - \dot{Z}^2 - Z\ddot{Z}}{(1 - \dot{Z}^2)^{3/2}} \left( \frac{1 - \dot{Z}^2}{Z^3} 2 f + \frac{\dot{Z}}{Z^2} 2 \partial_t f \right) + \frac{1 - \dot{Z}^2}{Z^2} W^{(1)}.
\end{equation}
Evaluating $xx$-component of the junction condition reads
\begin{align}
&K_{xx} -h_{xx}K = \frac{1 - \dot{Z}^2 - Z\ddot{Z}}{Z^2(1 - \dot{Z}^2)^{3/2}} + \frac{1}{Z^3 (1- \dot{Z}^2)^{5/2}}  \nonumber\\
&\times  \left[ -2(1 - \dot{Z}^2)^2 f + Z\dot{Z}(1 - \dot{Z}^2) \partial_t f - Z^2(1 - \dot{Z}^2) \partial_t^2 f -3 Z^2 \dot{Z} \ddot{Z} \partial_t f + Z \ddot{Z} (1 - \dot{Z}^2) f  \right],
\end{align}
\begin{equation}
T_{xx} = - \frac{2}{Z^3}\frac{1 - \dot{Z}^2 - Z\ddot{Z}}{(1 - \dot{Z}^2)^{3/2}} f - \frac{1}{Z^2} W^{(1)} .
\end{equation}
Combining the $tt$ and $xx$-component, we obtain
\begin{align}
&\frac{1}{Z^3\sqrt{1-\dot{Z}^2}} \left[ 2(1 - \dot{Z}^2)f + Z\dot{Z} \partial_t f - Z^2(1 - \dot{Z}^2) \partial_x^2 f \right] = 4 \dot{\bar{\phi}} \partial_t \varphi + \frac{1 - \dot{Z}^2 - Z\ddot{Z}}{(1 - \dot{Z}^2)^{3/2}} \left( \frac{1 - \dot{Z}^2}{Z^3} 2 f + \frac{\dot{Z}}{Z^2} 2 \partial_t f \right) \nonumber\\
& - \frac{1}{Z^3 (1- \dot{Z}^2)^{3/2}} \left[ -2(1 - \dot{Z}^2)^2 f + Z\dot{Z}(1 - \dot{Z}^2) \partial_t f - Z^2(1 - \dot{Z}^2) \partial_t^2 f -3 Z^2 \dot{Z} \ddot{Z} \partial_t f + Z \ddot{Z} (1 - \dot{Z}^2) f  \right]
\end{align}

Then, we evaluate the cross term, $tx$-component
\begin{equation}
K_{tx} -h_{tx}K = \frac{1}{Z^2 (1 - \dot{Z}^2)^{3/2}} \left[ \dot{Z} (1 - \dot{Z}^2 - Z \ddot{Z}) \partial_x f - Z(1 - \dot{Z}^2) \partial_x \partial_t f \right],
\end{equation}
\begin{equation}
\frac{1}{Z^2 (1 - \dot{Z}^2)^{3/2}} \left[ \dot{Z} (1 - \dot{Z}^2 - Z \ddot{Z}) \partial_x f - Z(1 - \dot{Z}^2) \partial_x \partial_t f \right] = 2 \dot{\bar{\phi}} \partial_x \varphi + \frac{\dot{Z}}{Z^2}  \frac{1 - \dot{Z}^2 - Z\ddot{Z}}{(1 - \dot{Z}^2)^{3/2}} \partial_x f.
\end{equation}
After some algebra, we obtain the following relation between metric and scalar perturbation
\begin{equation}
\varphi = - \frac{\partial_t f}{2 \dot{\bar{\phi}} Z \sqrt{1 - \dot{Z}^2}}.
\end{equation}
Combining this with the previous equation, we find
\begin{equation}
4 \dot{\bar{\phi}} \partial_t \varphi = - \frac{2 \partial_t^2 f}{Z \sqrt{1 - \dot{Z}^2}} + \frac{2 \partial_t f}{\dot{\bar{\phi}}^2 Z^2 (1 - \dot{Z}^2)} \left[ ( \ddot{\bar{\phi}} Z + \dot{\bar{\phi}} \dot{Z} ) \sqrt{1 - \dot{Z}^2} - \frac{\dot{\bar{\phi}}Z \dot{Z} \ddot{Z} }{\sqrt{1 - \dot{Z}^2}} \right].
\end{equation}

For a slow-roll brane that is discussed in section~\ref{sec:Inflation}, the profile is given by
\begin{equation}
Z \simeq \kappa t , (\dot{\bar\phi} \kappa t)^2 = \frac{\epsilon_1}{2 \sqrt{1 - \kappa^2}} = \mathrm{constant},
\end{equation}
where $\kappa = -1/\sqrt{2}$. Thus, the relation between metric and scalar perturbation is
\begin{equation}
\varphi(t,x) = - \frac{1}{2 \dot{\bar{\phi}} \kappa t \sqrt{1 - \kappa^2}} \partial_t f(t,x).
\end{equation}
Then, we find the following equation of motion for the metric perturbation
\begin{equation}
0 = - t^2 \partial_t^2 f +  ( 1- \kappa^2 ) t^2 \partial_x^2 f.
\end{equation}
Transforming to momentum space, the equation becomes
\begin{equation}
0 = f_p'' +  ( 1- \kappa^2 ) p^2 f_p,
\end{equation}
which yields the following solution
\begin{equation}
f_p(t) = c_1 e^{i \sqrt{1-\kappa^2} p t} + c_2 e^{-i \sqrt{1-\kappa^2} p t}.
\end{equation}
The solution implies that the slow-roll brane is stable under a small perturbation.


\section{Conclusion and discussion}
In this paper, we studied the effective theory on a two-dimensional end-of-the-world (EOW) brane embedded in an AdS$_3$ background and explored its applications to cosmology, as illustrated in Fig.~\ref{fig:inflationBCFT}. By introducing a scalar field localized on the brane, we obtained a rich class of brane dynamics that admits a natural cosmological interpretation.

To derive the effective theory, we integrated out the AdS$_3$ bulk and treated the embedding coordinates of the EOW brane as dynamical variables. A subtle point is that the equations of motion derived from the gauge-fixed effective action are not, by themselves, sufficient to reproduce the full Neumann boundary condition, which constitutes the fundamental equation of motion of the original brane model. The missing equations must instead be imposed as additional constraints. We clarified their origin by formulating the theory before fixing the worldvolume diffeomorphisms. In particular, within the minisuperspace ansatz, imposing the static gauge only after variation shows that the missing relation is precisely the Hamiltonian constraint.

We then considered the regime in which the gradients of the brane profile are small. In this limit, the effective brane action reduces to a Liouville-like theory coupled to the localized scalar field. Because the worldvolume diffeomorphisms have been gauge-fixed, the Liouville-like equations of motion must again be supplemented by the corresponding constraint equations in order to recover the full Neumann condition. By focusing on small perturbations around a dS brane solution, we worked out a covariant form of the effective action, where all equations of the original Neumann boundary condition can be derived from its equation of motion. This provides a direct derivation of the two-dimensional gravitational dynamics emerging on the EOW brane after the AdS$_3$ bulk is integrated out.

In the latter part of this paper, we investigated cosmological applications of the brane model. We first constructed a solution of the Neumann boundary condition whose induced geometry describes slow-roll inflation. From the brane trajectory, we reconstructed the behavior of the localized scalar field and the corresponding potential. Their dependence on the slow-roll parameter is consistent with the standard slow-roll picture, allowing the scalar field localized on the brane to be interpreted as an inflaton driving the cosmological expansion.

We next studied the Hartle--Hawking no-boundary description of the brane universe. In this framework, the semiclassical wavefunction is determined by a Euclidean path integral over geometries that are regular at the origin, without introducing an initial boundary there. Motivated by this prescription, we constructed both de Sitter and inflationary brane configurations in global AdS coordinates. These solutions originate from a smooth Euclidean geometry, admit a smooth analytic continuation to Lorentzian signature at a finite Euclidean time, and subsequently describe an expanding Lorentzian universe. For the de Sitter solution, we explicitly evaluated the on-shell action and obtained the probability for the nucleation of the brane universe. Moreover, in contrast to our earlier work \cite{Fujiki:2025yyf} where various cosmological solutions with initial singularities were found, here we were able to obtain an inflationary solution which is smoothly nucleated by an Euclidean instanton. 

Finally, we analyzed linear perturbations around the slow-roll brane solution. The linearized Neumann boundary conditions relate the perturbation of the localized scalar field to the deformation of the brane geometry and lead to an oscillatory equation for the physical brane perturbation. In particular, we found no exponentially growing mode within the regime of our approximation, indicating that the slow-roll brane solution is stable against small perturbations.

There are several interesting directions for future work. An important extension is to construct the corresponding effective theory for higher-dimensional EOW branes, which would allow us to discuss cosmologies closer to our four-dimensional universe. It would also be valuable to investigate quantum and higher-derivative corrections to the brane effective action and to clarify what they imply for quantum-gravitational effects in cosmology. Finally, because an EOW brane embedded in AdS is holographically dual to a boundary conformal field theory (BCFT), it would be particularly interesting to develop a field-theoretic understanding of the cosmological brane dynamics, including its semiclassical wavefunction and perturbations. Such a description may provide a useful framework for studying cosmology and quantum gravity from the viewpoint of holography.



\subsection*{Acknowledgments}
We would like to thank Liang Li for joining in this collaboration at an early stage. This work is supported by MEXT KAKENHI Grant-in-Aid for Transformative Research Areas (A) through the ``Extreme Universe'' collaboration: Grant Number 21H05187. TT is also supported by Inamori Research Institute for Science and by JSPS Grant-in-Aid for Scientific Research (B) No.~25K01000. 
KF is supported by Grant-in-Aid for JSPS Fellows No.~26KJ1554. 
M.K is  supported by Grant-in-Aid for JSPS Fellows No.~26KJ1545. 
Y.S is supported by RIKEN special postdoctoral researcher program.

\appendix

\section{Details of the gauge-unfixed minisuperspace reduction}
\label{app:unfixed-reduction-details}

We derive the geometric quantities used in
Section~\ref{sec:gauge-unfixed-minisuperspace}.  The Poincar\'e AdS$_3$ metric
is
\begin{equation}
 g_{\mu\nu}=\frac{1}{z^2}\operatorname{diag}(-1,1,1),
 \quad X^\mu=(t,z,x).
 \label{eq:appendix-poincare-metric}
\end{equation}
The nonvanishing Christoffel symbols relevant to the homogeneous embedding are
\begin{align}
 \Gamma^t{}_{tz}=\Gamma^t{}_{zt}&=-\frac{1}{z},
 &\Gamma^z{}_{tt}&=-\frac{1}{z},
 &\Gamma^z{}_{zz}&=-\frac{1}{z},
 \nonumber\\
 \Gamma^x{}_{xz}=\Gamma^x{}_{zx}&=-\frac{1}{z},
 &\Gamma^z{}_{xx}&=\frac{1}{z}.
 \label{eq:appendix-christoffels}
\end{align}
For
\begin{equation}
 X^\mu(\sigma,x)=\bigl(t(\sigma),Z(\sigma),x\bigr),
 \label{eq:appendix-mini-embedding}
\end{equation}
the tangent vectors are
\begin{equation}
 e_\sigma{}^\mu=(t',Z',0),
 \quad
 e_x{}^\mu=(0,0,1).
 \label{eq:appendix-mini-tangents}
\end{equation}
A spacelike unit normal one-form with the Type II orientation is
\begin{equation}
 n_\mu=
 \frac{1}{Z\sqrt{t'^{\,2}-Z'^{\,2}}}
 \bigl(-Z',t',0\bigr).
 \label{eq:appendix-mini-normal}
\end{equation}
It satisfies
\begin{equation}
 n_\mu e_a{}^\mu=0,
 \quad
 g^{\mu\nu}n_\mu n_\nu=1.
 \label{eq:appendix-normal-check}
\end{equation}
We use
\begin{equation}
 K_{ab}=n_\mu
 \left(
 \partial_a\partial_bX^\mu
 +\Gamma^\mu{}_{\rho\lambda}
 \partial_aX^\rho\partial_bX^\lambda
 \right).
 \label{eq:appendix-K-definition}
\end{equation}
The $\sigma\sigma$ component is
\begin{align}
 K_{\sigma\sigma}
 =&\frac{-Z'}{Z\sqrt{t'^{\,2}-Z'^{\,2}}}
 \left(t''-\frac{2t'Z'}{Z}\right)
 \nonumber\\
 &+\frac{t'}{Z\sqrt{t'^{\,2}-Z'^{\,2}}}
 \left(
 Z''-\frac{t'^{\,2}+Z'^{\,2}}{Z}
 \right)
 \nonumber\\
 =&\frac{Zt'Z''-ZZ't''-t'\bigl(t'^{\,2}-Z'^{\,2}\bigr)}
 {Z^2\sqrt{t'^{\,2}-Z'^{\,2}}}.
 \label{eq:appendix-Kss}
\end{align}
The spatial component is
\begin{equation}
 K_{xx}=n_z\Gamma^z{}_{xx}
 =\frac{t'}{Z^2\sqrt{t'^{\,2}-Z'^{\,2}}}.
 \label{eq:appendix-Kxx}
\end{equation}
Using
\begin{equation}
 h^{\sigma\sigma}=-\frac{Z^2}{t'^{\,2}-Z'^{\,2}},
 \quad
 h^{xx}=Z^2,
 \label{eq:appendix-induced-inverse-repeat}
\end{equation}
we obtain
\begin{align}
 K
 &=h^{\sigma\sigma}K_{\sigma\sigma}+h^{xx}K_{xx}
 \nonumber\\
 &=\frac{-Zt'Z''+Zt''Z'+2t'\bigl(t'^{\,2}-Z'^{\,2}\bigr)}
 {\bigl(t'^{\,2}-Z'^{\,2}\bigr)^{3/2}},
 \label{eq:appendix-K-trace}
\end{align}
which proves \eqref{eq:unfixed-K}.  Multiplying by the volume density gives
\begin{equation}
 \sqrt{-h}\,K
 =\frac{2t'}{Z^2}
 +\frac{t''Z'-t'Z''}{Z\bigl(t'^{\,2}-Z'^{\,2}\bigr)}.
 \label{eq:appendix-sqrth-K}
\end{equation}
The scalar and potential densities are
\begin{align}
 -\sqrt{-h}\,h^{ab}\partial_a\phi\partial_b\phi
 &=\frac{\phi'^{\,2}}{\sqrt{t'^{\,2}-Z'^{\,2}}},
 \label{eq:appendix-scalar-density}\\
 -\sqrt{-h}\,V(\phi)
 &=-\frac{\sqrt{t'^{\,2}-Z'^{\,2}}}{Z^2}V(\phi).
 \label{eq:appendix-potential-density}
\end{align}
Equations~\eqref{eq:appendix-sqrth-K}--\eqref{eq:appendix-potential-density}
lead directly to \eqref{eq:unfixed-brane-action}.

\section{Direct computation of $K_{tt}$, $K_{tx}$, and $K_{xx}$}
\label{app:direct-K-components}

In this appendix we compute the extrinsic curvature components of the
 brane directly in the AdS$_3$ metric. We fix the gauge, but we consider general brane profile $Z=Z(t,x)$.

The AdS$_3$ metric is
\begin{equation}
ds^2
=
\frac{dZ^2-dT^2+dX^2}{Z^2}.
\end{equation}
We use the component order
\begin{equation}
(T,Z,X).
\end{equation}
Thus
\begin{equation}
g_{\mu\nu}
=
\frac{1}{Z^2}
\mathrm{diag}(-1,+1,+1),
\quad
g^{\mu\nu}
=
Z^2
\mathrm{diag}(-1,+1,+1).
\end{equation}

The brane is written in a gauge-fixed form as
\begin{equation}
T=t,
\quad
X=x,
\quad
Z=Z(t,x).
\end{equation}
Equivalently, it is the level set
\begin{equation}
F(T,Z,X)=Z-Z(T,X)=0.
\end{equation}
We take the bulk region to be
\begin{equation}
F\geq 0,
\quad
\text{i.e.}
\quad
z\geq Z(T,X).
\end{equation}
The outward direction is therefore toward decreasing $F$. Since
\begin{equation}
dF=dZ-Z_t\,dT-Z_x\,dX,
\end{equation}
the outward-pointing unit normal one-form is
\begin{equation}
n_\mu dX^\mu
=
-\frac{dF}
{\sqrt{g^{\alpha\beta}\partial_\alpha F\partial_\beta F}}
=
\frac{-dZ+Z_t\,dT+Z_x\,dX}
{Z\sqrt{1-Z_t^2+Z_x^2}}.
\end{equation}
Therefore
\begin{equation}
n_\mu
=
\frac{1}
{Z\sqrt{1-Z_t^2+Z_x^2}}
\left(
Z_t,\,-1,\,Z_x
\right).
\end{equation}
Raising the index gives
\begin{equation}
n^\mu
=
\frac{Z}
{\sqrt{1-Z_t^2+Z_x^2}}
\left(
-Z_t,\,-1,\,Z_x
\right).
\end{equation}
The tangent vectors are
\begin{equation}
E_t^\mu=(1,Z_t,0),
\quad
E_x^\mu=(0,Z_x,1).
\end{equation}
They obey
\begin{equation}
n_\mu E_t^\mu=0,
\quad
n_\mu E_x^\mu=0,
\quad
n_\mu n^\mu=1.
\end{equation}

For notational economy in the intermediate steps, define
\begin{equation}
\mathcal A
\equiv
1-Z_t^2+Z_x^2.
\end{equation}
At the end we substitute $\mathcal A=1-Z_t^2+Z_x^2$.

The components of the normal one-form are
\begin{equation}
n_T=\frac{Z_t}{Z\sqrt{\mathcal A}},
\quad
n_Z=-\frac{1}{Z\sqrt{\mathcal A}},
\quad
n_X=\frac{Z_x}{Z\sqrt{\mathcal A}}.
\end{equation}
Here $Z_t$, $Z_x$, and $\mathcal A$ are functions of $T$ and $X$,
but not of the ambient coordinate $Z$. Hence
\begin{equation}
\partial_Z Z_t=0,
\quad
\partial_Z Z_x=0,
\quad
\partial_Z \mathcal A=0.
\end{equation}
We also have
\begin{equation}
\partial_T\mathcal A
=
-2Z_tZ_{tt}+2Z_xZ_{tx},
\end{equation}
and
\begin{equation}
\partial_X\mathcal A
=
-2Z_tZ_{tx}+2Z_xZ_{xx}.
\end{equation}

The derivatives of the normal components that will be needed below are
as follows:
\begin{equation}
\partial_T n_T
=
\frac{Z_{tt}}{Z\sqrt{\mathcal A}}
+
\frac{Z_t^2Z_{tt}-Z_tZ_xZ_{tx}}
{Z\mathcal A^{3/2}},
\end{equation}
\begin{equation}
\partial_X n_T
=
\frac{Z_{tx}}{Z\sqrt{\mathcal A}}
+
\frac{Z_t^2Z_{tx}-Z_tZ_xZ_{xx}}
{Z\mathcal A^{3/2}},
\end{equation}
\begin{equation}
\partial_Z n_T
=
-\frac{Z_t}{Z^2\sqrt{\mathcal A}},
\end{equation}
\begin{equation}
\partial_T n_Z
=
\frac{-Z_tZ_{tt}+Z_xZ_{tx}}
{Z\mathcal A^{3/2}},
\end{equation}
\begin{equation}
\partial_X n_Z
=
\frac{-Z_tZ_{tx}+Z_xZ_{xx}}
{Z\mathcal A^{3/2}},
\end{equation}
\begin{equation}
\partial_Z n_Z
=
\frac{1}{Z^2\sqrt{\mathcal A}},
\end{equation}
\begin{equation}
\partial_T n_X
=
\frac{Z_{tx}}{Z\sqrt{\mathcal A}}
+
\frac{Z_tZ_xZ_{tt}-Z_x^2Z_{tx}}
{Z\mathcal A^{3/2}},
\end{equation}
\begin{equation}
\partial_X n_X
=
\frac{Z_{xx}}{Z\sqrt{\mathcal A}}
+
\frac{Z_tZ_xZ_{tx}-Z_x^2Z_{xx}}
{Z\mathcal A^{3/2}},
\end{equation}
and
\begin{equation}
\partial_Z n_X
=
-\frac{Z_x}{Z^2\sqrt{\mathcal A}}.
\end{equation}

We use the convention
\begin{equation}
K_{ab}
=
E_a^\mu E_b^\nu \nabla_\mu n_\nu
=
E_a^\mu E_b^\nu
\left(
\partial_\mu n_\nu
-
\Gamma^\rho_{\mu\nu}n_\rho
\right).
\end{equation}
The nonzero Christoffel symbols of the AdS$_3$ metric are
\begin{equation}
\Gamma^T_{TZ}
=
\Gamma^T_{ZT}
=
-\frac{1}{Z},
\quad
\Gamma^X_{XZ}
=
\Gamma^X_{ZX}
=
-\frac{1}{Z},
\end{equation}
and
\begin{equation}
\Gamma^Z_{TT}
=
-\frac{1}{Z},
\quad
\Gamma^Z_{ZZ}
=
-\frac{1}{Z},
\quad
\Gamma^Z_{XX}
=
\frac{1}{Z}.
\end{equation}

\subsection*{The $tt$ component}

First compute the derivative part
\begin{equation}
\mathcal D_{tt}
\equiv
E_t^\mu E_t^\nu \partial_\mu n_\nu.
\end{equation}
Since
\begin{equation}
E_t^\mu=(1,Z_t,0),
\end{equation}
we have
\begin{equation}
\mathcal D_{tt}
=
\partial_T n_T
+
Z_t\partial_Z n_T
+
Z_t\partial_T n_Z
+
Z_t^2\partial_Z n_Z.
\end{equation}
Substituting the derivatives above,
\begin{equation}
\begin{aligned}
\partial_T n_T+Z_t\partial_T n_Z
&=
\frac{Z_{tt}}{Z\sqrt{\mathcal A}}
+
\frac{Z_t^2Z_{tt}-Z_tZ_xZ_{tx}}
{Z\mathcal A^{3/2}}
+
\frac{-Z_t^2Z_{tt}+Z_tZ_xZ_{tx}}
{Z\mathcal A^{3/2}}
\\
&=
\frac{Z_{tt}}{Z\sqrt{\mathcal A}}.
\end{aligned}
\end{equation}
Thus the $Z_{tx}$ terms cancel. The ambient $Z$-derivative terms also
cancel:
\begin{equation}
Z_t\partial_Z n_T+Z_t^2\partial_Z n_Z
=
-\frac{Z_t^2}{Z^2\sqrt{\mathcal A}}
+
\frac{Z_t^2}{Z^2\sqrt{\mathcal A}}
=
0.
\end{equation}
Therefore
\begin{equation}
\mathcal D_{tt}
=
\frac{Z_{tt}}{Z\sqrt{\mathcal A}}.
\end{equation}

Next compute the connection part
\begin{equation}
\mathcal G_{tt}
\equiv
E_t^\mu E_t^\nu\Gamma^\rho_{\mu\nu}n_\rho.
\end{equation}
The relevant contraction is
\begin{equation}
\Gamma^\rho_{\mu\nu}E_t^\mu E_t^\nu
=
\left(
-\frac{2Z_t}{Z},
-\frac{1+Z_t^2}{Z},
0
\right).
\end{equation}
Hence
\begin{equation}
\begin{aligned}
\mathcal G_{tt}
&=
\left(-\frac{2Z_t}{Z}\right)n_T
+
\left(-\frac{1+Z_t^2}{Z}\right)n_Z
\\
&=
-\frac{2Z_t^2}{Z^2\sqrt{\mathcal A}}
+
\frac{1+Z_t^2}{Z^2\sqrt{\mathcal A}}
\\
&=
\frac{1-Z_t^2}{Z^2\sqrt{\mathcal A}}.
\end{aligned}
\end{equation}
Therefore
\begin{equation}
K_{tt}
=
\mathcal D_{tt}-\mathcal G_{tt}
=
\frac{Z_{tt}}{Z\sqrt{\mathcal A}}
-
\frac{1-Z_t^2}{Z^2\sqrt{\mathcal A}}.
\end{equation}
Equivalently,
\begin{equation}
K_{tt}
=
\frac{ZZ_{tt}-1+Z_t^2}
{Z^2\sqrt{\mathcal A}}.
\end{equation}
Restoring $\mathcal A=1-Z_t^2+Z_x^2$, we obtain
\begin{equation}
K_{tt}
=
\frac{ZZ_{tt}-1+Z_t^2}
{Z^2\sqrt{1-Z_t^2+Z_x^2}}
.
\end{equation}

\subsection*{The $tx$ component}

Next compute
\begin{equation}
\mathcal D_{tx}
\equiv
E_t^\mu E_x^\nu\partial_\mu n_\nu.
\end{equation}
Using
\begin{equation}
E_t^\mu=(1,Z_t,0),
\quad
E_x^\nu=(0,Z_x,1),
\end{equation}
we get
\begin{equation}
\mathcal D_{tx}
=
Z_x\partial_T n_Z
+
\partial_T n_X
+
Z_tZ_x\partial_Z n_Z
+
Z_t\partial_Z n_X.
\end{equation}
First, the $T$-derivative terms give
\begin{equation}
\begin{aligned}
Z_x\partial_T n_Z+\partial_T n_X
&=
\frac{-Z_tZ_xZ_{tt}+Z_x^2Z_{tx}}
{Z\mathcal A^{3/2}}
+
\frac{Z_{tx}}{Z\sqrt{\mathcal A}}
+
\frac{Z_tZ_xZ_{tt}-Z_x^2Z_{tx}}
{Z\mathcal A^{3/2}}
\\
&=
\frac{Z_{tx}}{Z\sqrt{\mathcal A}}.
\end{aligned}
\end{equation}
Thus the $Z_{tt}$ terms cancel. The ambient $Z$-derivative terms are
\begin{equation}
Z_tZ_x\partial_Z n_Z
+
Z_t\partial_Z n_X
=
\frac{Z_tZ_x}{Z^2\sqrt{\mathcal A}}
-
\frac{Z_tZ_x}{Z^2\sqrt{\mathcal A}}
=
0.
\end{equation}
Therefore
\begin{equation}
\mathcal D_{tx}
=
\frac{Z_{tx}}{Z\sqrt{\mathcal A}}.
\end{equation}

The connection part is
\begin{equation}
\mathcal G_{tx}
\equiv
E_t^\mu E_x^\nu\Gamma^\rho_{\mu\nu}n_\rho.
\end{equation}
The relevant contraction is
\begin{equation}
\Gamma^\rho_{\mu\nu}E_t^\mu E_x^\nu
=
\left(
-\frac{Z_x}{Z},
-\frac{Z_tZ_x}{Z},
-\frac{Z_t}{Z}
\right).
\end{equation}
Thus
\begin{equation}
\begin{aligned}
\mathcal G_{tx}
&=
\left(-\frac{Z_x}{Z}\right)n_T
+
\left(-\frac{Z_tZ_x}{Z}\right)n_Z
+
\left(-\frac{Z_t}{Z}\right)n_X
\\
&=
-\frac{Z_tZ_x}{Z^2\sqrt{\mathcal A}}
+
\frac{Z_tZ_x}{Z^2\sqrt{\mathcal A}}
-
\frac{Z_tZ_x}{Z^2\sqrt{\mathcal A}}
\\
&=
-\frac{Z_tZ_x}{Z^2\sqrt{\mathcal A}}.
\end{aligned}
\end{equation}
Therefore
\begin{equation}
K_{tx}
=
\mathcal D_{tx}-\mathcal G_{tx}
=
\frac{Z_{tx}}{Z\sqrt{\mathcal A}}
+
\frac{Z_tZ_x}{Z^2\sqrt{\mathcal A}}.
\end{equation}
Equivalently,
\begin{equation}
K_{tx}
=
\frac{ZZ_{tx}+Z_tZ_x}
{Z^2\sqrt{\mathcal A}}.
\end{equation}
Restoring $\mathcal A=1-Z_t^2+Z_x^2$, we obtain
\begin{equation}
K_{tx}
=
\frac{ZZ_{tx}+Z_tZ_x}
{Z^2\sqrt{1-Z_t^2+Z_x^2}}
.
\end{equation}

As a check, one may compute $K_{xt}$ instead. The derivative part is
\begin{equation}
E_x^\mu E_t^\nu\partial_\mu n_\nu
=
\partial_X n_T
+
Z_x\partial_Z n_T
+
Z_t\partial_X n_Z
+
Z_tZ_x\partial_Z n_Z.
\end{equation}
Using the derivatives above,
\begin{equation}
\partial_X n_T+Z_t\partial_X n_Z
=
\frac{Z_{tx}}{Z\sqrt{\mathcal A}},
\end{equation}
and
\begin{equation}
Z_x\partial_Z n_T+Z_tZ_x\partial_Z n_Z=0.
\end{equation}
Therefore
\begin{equation}
E_x^\mu E_t^\nu\partial_\mu n_\nu
=
\frac{Z_{tx}}{Z\sqrt{\mathcal A}}.
\end{equation}
The connection part is symmetric, so $K_{xt}=K_{tx}$, as required.

\subsection*{The $xx$ component}

Finally compute
\begin{equation}
\mathcal D_{xx}
\equiv
E_x^\mu E_x^\nu\partial_\mu n_\nu.
\end{equation}
Since
\begin{equation}
E_x^\mu=(0,Z_x,1),
\end{equation}
we have
\begin{equation}
\mathcal D_{xx}
=
Z_x\partial_X n_Z
+
\partial_X n_X
+
Z_x^2\partial_Z n_Z
+
Z_x\partial_Z n_X.
\end{equation}
First, the $X$-derivative terms give
\begin{equation}
\begin{aligned}
Z_x\partial_X n_Z+\partial_X n_X
&=
\frac{-Z_tZ_xZ_{tx}+Z_x^2Z_{xx}}
{Z\mathcal A^{3/2}}
+
\frac{Z_{xx}}{Z\sqrt{\mathcal A}}
+
\frac{Z_tZ_xZ_{tx}-Z_x^2Z_{xx}}
{Z\mathcal A^{3/2}}
\\
&=
\frac{Z_{xx}}{Z\sqrt{\mathcal A}}.
\end{aligned}
\end{equation}
Thus the $Z_{tx}$ terms cancel. The ambient $Z$-derivative terms are
\begin{equation}
Z_x^2\partial_Z n_Z
+
Z_x\partial_Z n_X
=
\frac{Z_x^2}{Z^2\sqrt{\mathcal A}}
-
\frac{Z_x^2}{Z^2\sqrt{\mathcal A}}
=
0.
\end{equation}
Therefore
\begin{equation}
\mathcal D_{xx}
=
\frac{Z_{xx}}{Z\sqrt{\mathcal A}}.
\end{equation}

The connection part is
\begin{equation}
\mathcal G_{xx}
\equiv
E_x^\mu E_x^\nu\Gamma^\rho_{\mu\nu}n_\rho.
\end{equation}
The relevant contraction is
\begin{equation}
\Gamma^\rho_{\mu\nu}E_x^\mu E_x^\nu
=
\left(
0,
\frac{1-Z_x^2}{Z},
-\frac{2Z_x}{Z}
\right).
\end{equation}
Thus
\begin{equation}
\begin{aligned}
\mathcal G_{xx}
&=
\left(\frac{1-Z_x^2}{Z}\right)n_Z
+
\left(-\frac{2Z_x}{Z}\right)n_X
\\
&=
-\frac{1-Z_x^2}{Z^2\sqrt{\mathcal A}}
-
\frac{2Z_x^2}{Z^2\sqrt{\mathcal A}}
\\
&=
-\frac{1+Z_x^2}{Z^2\sqrt{\mathcal A}}.
\end{aligned}
\end{equation}
Therefore
\begin{equation}
K_{xx}
=
\mathcal D_{xx}-\mathcal G_{xx}
=
\frac{Z_{xx}}{Z\sqrt{\mathcal A}}
+
\frac{1+Z_x^2}{Z^2\sqrt{\mathcal A}}.
\end{equation}
Equivalently,
\begin{equation}
K_{xx}
=
\frac{ZZ_{xx}+1+Z_x^2}
{Z^2\sqrt{\mathcal A}}.
\end{equation}
Restoring $\mathcal A=1-Z_t^2+Z_x^2$, we obtain
\begin{equation}
K_{xx}
=
\frac{ZZ_{xx}+1+Z_x^2}
{Z^2\sqrt{1-Z_t^2+Z_x^2}}
.
\end{equation}

\subsection*{Trace}

The induced metric and its inverse are
\begin{equation}
h_{ab}
=
\frac{1}{Z^2}
\begin{pmatrix}
-1+Z_t^2 & Z_tZ_x \\
Z_tZ_x & 1+Z_x^2
\end{pmatrix},
\end{equation}
and
\begin{equation}
h^{ab}
=
\frac{Z^2}{1-Z_t^2+Z_x^2}
\begin{pmatrix}
-(1+Z_x^2) & Z_tZ_x \\
Z_tZ_x & 1-Z_t^2
\end{pmatrix}.
\end{equation}
Thus
\begin{equation}
K
=
h^{tt}K_{tt}
+
2h^{tx}K_{tx}
+
h^{xx}K_{xx}.
\end{equation}
Substituting the components above gives
\begin{equation}
\begin{aligned}
K
=
\frac{1}{(1-Z_t^2+Z_x^2)^{3/2}}
\Big[
&
-(1+Z_x^2)(ZZ_{tt}-1+Z_t^2)
\\
&
+2Z_tZ_x(ZZ_{tx}+Z_tZ_x)
\\
&
+(1-Z_t^2)(ZZ_{xx}+1+Z_x^2)
\Big].
\end{aligned}
\end{equation}
Expanding the numerator, this becomes
\begin{equation}
K
=
\frac{
2(1-Z_t^2+Z_x^2)
-
Z\left[
(1+Z_x^2)Z_{tt}
+
(Z_t^2-1)Z_{xx}
-
2Z_tZ_xZ_{tx}
\right]
}
{
(1-Z_t^2+Z_x^2)^{3/2}
}.
\end{equation}
Finally,
\begin{equation}
\sqrt{-h}
=
\frac{\sqrt{1-Z_t^2+Z_x^2}}{Z^2},
\end{equation}
so
\begin{equation}
\sqrt{-h}\,K
=
\frac{
2(1-Z_t^2+Z_x^2)
-
Z\left[
(1+Z_x^2)Z_{tt}
+
(Z_t^2-1)Z_{xx}
-
2Z_tZ_xZ_{tx}
\right]
}
{
Z^2(1-Z_t^2+Z_x^2)
}.
\end{equation}

For a constant-$Z$ brane,
\begin{equation}
Z_t=Z_x=Z_{tt}=Z_{tx}=Z_{xx}=0,
\end{equation}
we find
\begin{equation}
K_{tt}=-\frac{1}{Z^2},
\quad
K_{tx}=0,
\quad
K_{xx}=\frac{1}{Z^2},
\quad
K=2.
\end{equation}
This agrees with the Type-II orientation used in the main text. If the
opposite bulk region is chosen, the outward normal is reversed and
\begin{equation}
K_{ab}\rightarrow -K_{ab},
\quad
K\rightarrow -K.
\end{equation}
\eject

 \bibliography{braneinfla}

@article{Randall:1999ee,
    author = "Randall, Lisa and Sundrum, Raman",
    title = "{A Large mass hierarchy from a small extra dimension}",
    eprint = "hep-ph/9905221",
    archivePrefix = "arXiv",
    reportNumber = "MIT-CTP-2860, PUPT-1860, BUHEP-99-9",
    doi = "10.1103/PhysRevLett.83.3370",
    journal = "Phys. Rev. Lett.",
    volume = "83",
    pages = "3370--3373",
    year = "1999"
}

@article{Randall:1999vf,
    author = "Randall, Lisa and Sundrum, Raman",
    title = "{An Alternative to compactification}",
    eprint = "hep-th/9906064",
    archivePrefix = "arXiv",
    reportNumber = "MIT-CTP-2874, PUPT-1867, BUHEP-99-13",
    doi = "10.1103/PhysRevLett.83.4690",
    journal = "Phys. Rev. Lett.",
    volume = "83",
    pages = "4690--4693",
    year = "1999"
}

@article{Garriga:1999yh,
    author = "Garriga, Jaume and Tanaka, Takahiro",
    title = "{Gravity in the brane world}",
    eprint = "hep-th/9911055",
    archivePrefix = "arXiv",
    reportNumber = "UAB-FT-476, OU-TAP-106",
    doi = "10.1103/PhysRevLett.84.2778",
    journal = "Phys. Rev. Lett.",
    volume = "84",
    pages = "2778--2781",
    year = "2000"
}

@article{Shiromizu:1999wj,
    author = "Shiromizu, Tetsuya and Maeda, Kei-ichi and Sasaki, Misao",
    title = "{The Einstein equation on the 3-brane world}",
    eprint = "gr-qc/9910076",
    archivePrefix = "arXiv",
    reportNumber = "DAMTP-1999-150, OUTAP-103, UTAP-349, RESCEU-40-99",
    doi = "10.1103/PhysRevD.62.024012",
    journal = "Phys. Rev. D",
    volume = "62",
    pages = "024012",
    year = "2000"
}

@article{Gubser:1999vj,
    author = "Gubser, Steven S.",
    title = "{AdS / CFT and gravity}",
    eprint = "hep-th/9912001",
    archivePrefix = "arXiv",
    reportNumber = "HUTP-99-A065",
    doi = "10.1103/PhysRevD.63.084017",
    journal = "Phys. Rev. D",
    volume = "63",
    pages = "084017",
    year = "2001"
}

@article{Karch:2000ct,
    author = "Karch, Andreas and Randall, Lisa",
    editor = "Duff, Michael J. and Liu, J. T. and Lu, J.",
    title = "{Locally localized gravity}",
    eprint = "hep-th/0011156",
    archivePrefix = "arXiv",
    reportNumber = "MIT-CTP-3099",
    doi = "10.1088/1126-6708/2001/05/008",
    journal = "JHEP",
    volume = "05",
    pages = "008",
    year = "2001"
}

@article{Shiromizu:2001jm,
    author = "Shiromizu, Tetsuya and Ida, Daisuke",
    title = "{Anti-de Sitter no hair, AdS / CFT and the brane world}",
    eprint = "hep-th/0102035",
    archivePrefix = "arXiv",
    reportNumber = "RESCEU-43-00, UTAP-384",
    doi = "10.1103/PhysRevD.64.044015",
    journal = "Phys. Rev. D",
    volume = "64",
    pages = "044015",
    year = "2001"
}

@article{Koyama:2001rf,
    author = "Koyama, Kazuya and Soda, Jiro",
    title = "{Strongly coupled CFT in FRW universe from AdS / CFT correspondence}",
    eprint = "hep-th/0101164",
    archivePrefix = "arXiv",
    doi = "10.1088/1126-6708/2001/05/027",
    journal = "JHEP",
    volume = "05",
    pages = "027",
    year = "2001"
}

@article{Shiromizu:2001ve,
    author = "Shiromizu, Tetsuya and Torii, Takashi and Ida, Daisuke",
    title = "{Brane world and holography}",
    eprint = "hep-th/0105256",
    archivePrefix = "arXiv",
    reportNumber = "UTAP-391, RESCEU-8-01",
    doi = "10.1088/1126-6708/2002/03/007",
    journal = "JHEP",
    volume = "03",
    pages = "007",
    year = "2002"
}

@article{Maldacena:1997re,
    author = "Maldacena, Juan Martin",
    title = "{The Large $N$ limit of superconformal field theories and supergravity}",
    eprint = "hep-th/9711200",
    archivePrefix = "arXiv",
    reportNumber = "HUTP-97-A097, HUTP-98-A097",
    doi = "10.4310/ATMP.1998.v2.n2.a1",
    journal = "Adv. Theor. Math. Phys.",
    volume = "2",
    pages = "231--252",
    year = "1998"
}

@article{Gubser:1998bc,
    author = "Gubser, S. S. and Klebanov, Igor R. and Polyakov, Alexander M.",
    title = "{Gauge theory correlators from noncritical string theory}",
    eprint = "hep-th/9802109",
    archivePrefix = "arXiv",
    reportNumber = "PUPT-1767",
    doi = "10.1016/S0370-2693(98)00377-3",
    journal = "Phys. Lett. B",
    volume = "428",
    pages = "105--114",
    year = "1998"
}

@article{Witten:1998qj,
    author = "Witten, Edward",
    title = "{Anti de Sitter space and holography}",
    eprint = "hep-th/9802150",
    archivePrefix = "arXiv",
    reportNumber = "IASSNS-HEP-98-15",
    doi = "10.4310/ATMP.1998.v2.n2.a2",
    journal = "Adv. Theor. Math. Phys.",
    volume = "2",
    pages = "253--291",
    year = "1998"
}

@article{Takayanagi:2011zk,
    author = "Takayanagi, Tadashi",
    title = "{Holographic Dual of BCFT}",
    eprint = "1105.5165",
    archivePrefix = "arXiv",
    primaryClass = "hep-th",
    reportNumber = "IPMU11-0091",
    doi = "10.1103/PhysRevLett.107.101602",
    journal = "Phys. Rev. Lett.",
    volume = "107",
    pages = "101602",
    year = "2011"
}

@article{Fujita:2011fp,
    author = "Fujita, Mitsutoshi and Takayanagi, Tadashi and Tonni, Erik",
    title = "{Aspects of AdS/BCFT}",
    eprint = "1108.5152",
    archivePrefix = "arXiv",
    primaryClass = "hep-th",
    reportNumber = "IPMU-11-0136, MIT-CTP-4289",
    doi = "10.1007/JHEP11(2011)043",
    journal = "JHEP",
    volume = "11",
    pages = "043",
    year = "2011"
}

@article{Cardy:1984bb,
    author = "Cardy, John L.",
    title = "{Conformal Invariance and Surface Critical Behavior}",
    doi = "10.1016/0550-3213(84)90241-4",
    journal = "Nucl. Phys. B",
    volume = "240",
    pages = "514--532",
    year = "1984"
}

@article{Cardy:2004hm,
    author = "Cardy, John L.",
    title = "{Boundary conformal field theory}",
    eprint = "hep-th/0411189",
    archivePrefix = "arXiv",
    month = "11",
    year = "2004"
}

@article{Cooper:2018cmb,
    author = "Cooper, Sean and Rozali, Moshe and Swingle, Brian and Van Raamsdonk, Mark and Waddell, Christopher and Wakeham, David",
    title = "{Black hole microstate cosmology}",
    eprint = "1810.10601",
    archivePrefix = "arXiv",
    primaryClass = "hep-th",
    doi = "10.1007/JHEP07(2019)065",
    journal = "JHEP",
    volume = "07",
    pages = "065",
    year = "2019"
}

@article{Antonini:2019qkt,
    author = "Antonini, Stefano and Swingle, Brian",
    title = "{Cosmology at the end of the world}",
    eprint = "1907.06667",
    archivePrefix = "arXiv",
    primaryClass = "hep-th",
    doi = "10.1038/s41567-020-0909-6",
    journal = "Nature Phys.",
    volume = "16",
    number = "8",
    pages = "881--886",
    year = "2020"
}

@article{VanRaamsdonk:2021qgv,
    author = "Van Raamsdonk, Mark",
    title = "{Cosmology from confinement?}",
    eprint = "2102.05057",
    archivePrefix = "arXiv",
    primaryClass = "hep-th",
    doi = "10.1007/JHEP03(2022)039",
    journal = "JHEP",
    volume = "03",
    pages = "039",
    year = "2022"
}

@article{Omiya:2021olc,
    author = "Omiya, Hidetoshi and Wei, Zixia",
    title = "{Causal structures and nonlocality in double holography}",
    eprint = "2107.01219",
    archivePrefix = "arXiv",
    primaryClass = "hep-th",
    reportNumber = "YITP-21-51",
    doi = "10.1007/JHEP07(2022)128",
    journal = "JHEP",
    volume = "07",
    pages = "128",
    year = "2022"
}

@article{Antonini:2022blk,
    author = "Antonini, Stefano and Simidzija, Petar and Swingle, Brian and Van Raamsdonk, Mark",
    title = "{Cosmology from the vacuum}",
    eprint = "2203.11220",
    archivePrefix = "arXiv",
    primaryClass = "hep-th",
    doi = "10.1088/1361-6382/ad1d46",
    journal = "Class. Quant. Grav.",
    volume = "41",
    number = "4",
    pages = "045008",
    year = "2024"
}

@article{Waddell:2022fbn,
    author = "Waddell, Chris",
    title = "{Bottom-up holographic models for cosmology}",
    eprint = "2203.03096",
    archivePrefix = "arXiv",
    primaryClass = "hep-th",
    doi = "10.1007/JHEP09(2022)176",
    journal = "JHEP",
    volume = "09",
    pages = "176",
    year = "2022"
}

@article{Fujiki:2025yyf,
    author = "Fujiki, Kosei and Kanda, Hiroki and Kohara, Michitaka and Takayanagi, Tadashi",
    title = "{Brane cosmology from AdS/BCFT}",
    eprint = "2501.05036",
    archivePrefix = "arXiv",
    primaryClass = "hep-th",
    reportNumber = "YITP-24-180",
    doi = "10.1007/JHEP03(2025)135",
    journal = "JHEP",
    volume = "03",
    pages = "135",
    year = "2025"
}

@article{Strominger:2001pn,
    author = "Strominger, Andrew",
    title = "{The dS / CFT correspondence}",
    eprint = "hep-th/0106113",
    archivePrefix = "arXiv",
    doi = "10.1088/1126-6708/2001/10/034",
    journal = "JHEP",
    volume = "10",
    pages = "034",
    year = "2001"
}

@inproceedings{Witten:2001kn,
    author = "Witten, Edward",
    title = "{Quantum gravity in de Sitter space}",
    booktitle = "{Strings 2001: International Conference}",
    eprint = "hep-th/0106109",
    archivePrefix = "arXiv",
    month = "6",
    year = "2001"
}

@inproceedings{Spradlin:2001pw,
    author = "Spradlin, Marcus and Strominger, Andrew and Volovich, Anastasia",
    title = "{Les Houches lectures on de Sitter space}",
    booktitle = "{Les Houches Summer School: Session 76: Euro Summer School on Unity of Fundamental Physics: Gravity, Gauge Theory and Strings}",
    eprint = "hep-th/0110007",
    archivePrefix = "arXiv",
    pages = "423--453",
    month = "10",
    year = "2001"
}

@article{Bousso:2001mw,
    author = "Bousso, Raphael and Maloney, Alexander and Strominger, Andrew",
    title = "{Conformal vacua and entropy in de Sitter space}",
    eprint = "hep-th/0112218",
    archivePrefix = "arXiv",
    doi = "10.1103/PhysRevD.65.104039",
    journal = "Phys. Rev. D",
    volume = "65",
    pages = "104039",
    year = "2002"
}

@article{Balasubramanian:2002zh,
    author = "Balasubramanian, Vijay and de Boer, Jan and Minic, Djordje",
    editor = "de Wit, B. and Vandoren, S.",
    title = "{Notes on de Sitter space and holography}",
    eprint = "hep-th/0207245",
    archivePrefix = "arXiv",
    reportNumber = "VPI-IPPAP-02-05, UPR-1008-T, IFTA-2002-26",
    doi = "10.1016/S0003-4916(02)00020-9",
    journal = "Class. Quant. Grav.",
    volume = "19",
    pages = "5655--5700",
    year = "2002"
}

@article{Klemm:2001ea,
    author = "Klemm, Dietmar",
    title = "{Some aspects of the de Sitter / CFT correspondence}",
    eprint = "hep-th/0106247",
    archivePrefix = "arXiv",
    reportNumber = "IFUM-690-FT",
    doi = "10.1016/S0550-3213(02)00007-X",
    journal = "Nucl. Phys. B",
    volume = "625",
    pages = "295--311",
    year = "2002"
}

@article{Balasubramanian:2001nb,
    author = "Balasubramanian, Vijay and de Boer, Jan and Minic, Djordje",
    title = "{Mass, entropy and holography in asymptotically de Sitter spaces}",
    eprint = "hep-th/0110108",
    archivePrefix = "arXiv",
    reportNumber = "VPI-IPPAP-01-01, UPR-964-T",
    doi = "10.1103/PhysRevD.65.123508",
    journal = "Phys. Rev. D",
    volume = "65",
    pages = "123508",
    year = "2002"
}

@article{Strominger:2001gp,
    author = "Strominger, Andrew",
    title = "{Inflation and the dS / CFT correspondence}",
    eprint = "hep-th/0110087",
    archivePrefix = "arXiv",
    doi = "10.1088/1126-6708/2001/11/049",
    journal = "JHEP",
    volume = "11",
    pages = "049",
    year = "2001"
}

@article{Maldacena:2002vr,
    author = "Maldacena, Juan Martin",
    title = "{Non-Gaussian features of primordial fluctuations in single field inflationary models}",
    eprint = "astro-ph/0210603",
    archivePrefix = "arXiv",
    doi = "10.1088/1126-6708/2003/05/013",
    journal = "JHEP",
    volume = "05",
    pages = "013",
    year = "2003"
}

@article{Larsen:2002et,
    author = "Larsen, Finn and van der Schaar, Jan Pieter and Leigh, Robert G.",
    title = "{De Sitter holography and the cosmic microwave background}",
    eprint = "hep-th/0202127",
    archivePrefix = "arXiv",
    reportNumber = "MCTP-02-09, ILL-TH-02-02",
    doi = "10.1088/1126-6708/2002/04/047",
    journal = "JHEP",
    volume = "04",
    pages = "047",
    year = "2002"
}

@article{Larsen:2003pf,
    author = "Larsen, Finn and McNees, Robert",
    title = "{Inflation and de Sitter holography}",
    eprint = "hep-th/0307026",
    archivePrefix = "arXiv",
    reportNumber = "MCTP-03-32",
    doi = "10.1088/1126-6708/2003/07/051",
    journal = "JHEP",
    volume = "07",
    pages = "051",
    year = "2003"
}

@article{McFadden:2009fg,
    author = "McFadden, Paul and Skenderis, Kostas",
    title = "{Holography for Cosmology}",
    eprint = "0907.5542",
    archivePrefix = "arXiv",
    primaryClass = "hep-th",
    reportNumber = "ITF-22",
    doi = "10.1103/PhysRevD.81.021301",
    journal = "Phys. Rev. D",
    volume = "81",
    pages = "021301",
    year = "2010"
}

@article{McFadden:2010na,
    author = "McFadden, Paul and Skenderis, Kostas",
    editor = "Basilakos, Spyros and Cadoni, Mariano and Cavaglia, Marco and Christodoulakis, Theodosios and Vagenas, Elias C.",
    title = "{The Holographic Universe}",
    eprint = "1001.2007",
    archivePrefix = "arXiv",
    primaryClass = "hep-th",
    reportNumber = "ITFA-10-02",
    doi = "10.1088/1742-6596/222/1/012007",
    journal = "J. Phys. Conf. Ser.",
    volume = "222",
    pages = "012007",
    year = "2010"
}

@article{Akal:2020wfl,
    author = "Akal, Ibrahim and Kusuki, Yuya and Takayanagi, Tadashi and Wei, Zixia",
    title = "{Codimension two holography for wedges}",
    eprint = "2007.06800",
    archivePrefix = "arXiv",
    primaryClass = "hep-th",
    reportNumber = "YITP-20-91, IPMU20-0079",
    doi = "10.1103/PhysRevD.102.126007",
    journal = "Phys. Rev. D",
    volume = "102",
    number = "12",
    pages = "126007",
    year = "2020"
}

@article{Chen:2020tes,
    author = "Chen, Yiming and Gorbenko, Victor and Maldacena, Juan",
    title = "{Bra-ket wormholes in gravitationally prepared states}",
    eprint = "2007.16091",
    archivePrefix = "arXiv",
    primaryClass = "hep-th",
    doi = "10.1007/JHEP02(2021)009",
    journal = "JHEP",
    volume = "02",
    pages = "009",
    year = "2021"
}

@article{Wei:2024zez,
    author = "Wei, Zixia",
    title = "{Holographic dual of crosscap conformal field theory}",
    eprint = "2405.03755",
    archivePrefix = "arXiv",
    primaryClass = "hep-th",
    doi = "10.1007/JHEP03(2025)086",
    journal = "JHEP",
    volume = "03",
    pages = "086",
    year = "2025"
}

@article{Hoback:2026yqj,
    author = "Hoback, Sarah and Jafferis, Daniel L. and Wei, Zixia",
    title = "{Monitoring a de Sitter universe through an anti-de Sitter window}",
    eprint = "2606.31705",
    archivePrefix = "arXiv",
    primaryClass = "hep-th",
    month = "6",
    year = "2026"
}

@article{Kanda:2023zse,
    author = "Kanda, Hiroki and Sato, Masahide and Suzuki, Yu-ki and Takayanagi, Tadashi and Wei, Zixia",
    title = "{AdS/BCFT with brane-localized scalar field}",
    eprint = "2302.03895",
    archivePrefix = "arXiv",
    primaryClass = "hep-th",
    reportNumber = "YITP-23-07",
    doi = "10.1007/JHEP03(2023)105",
    journal = "JHEP",
    volume = "03",
    pages = "105",
    year = "2023"
}

@article{Kanda:2023jyi,
    author = "Kanda, Hiroki and Kawamoto, Taishi and Suzuki, Yu-ki and Takayanagi, Tadashi and Tasuki, Kenya and Wei, Zixia",
    title = "{Entanglement phase transition in holographic pseudo entropy}",
    eprint = "2311.13201",
    archivePrefix = "arXiv",
    primaryClass = "hep-th",
    reportNumber = "YITP-23-148",
    doi = "10.1007/JHEP03(2024)060",
    journal = "JHEP",
    volume = "03",
    pages = "060",
    year = "2024"
}

@article{Maeda:2026awj,
    author = "Maeda, Ryota and Nakamura, Nanami and Takayanagi, Tadashi",
    title = "{Holographic Dual of PT Symmetric BCFT}",
    eprint = "2606.18629",
    archivePrefix = "arXiv",
    primaryClass = "hep-th",
    reportNumber = "YITP-26-68",
    month = "6",
    year = "2026"
}

@article{Kanno:2002iaa,
    author = "Kanno, Sugumi and Soda, Jiro",
    title = "{Brane world effective action at low-energies and AdS / CFT}",
    eprint = "hep-th/0205188",
    archivePrefix = "arXiv",
    reportNumber = "KUCP-0209",
    doi = "10.1103/PhysRevD.66.043526",
    journal = "Phys. Rev. D",
    volume = "66",
    pages = "043526",
    year = "2002"
}

@article{Suzuki:2022xwv,
    author = "Suzuki, Kenta and Takayanagi, Tadashi",
    title = "{BCFT and Islands in two dimensions}",
    eprint = "2202.08462",
    archivePrefix = "arXiv",
    primaryClass = "hep-th",
    reportNumber = "YITP-22-14, IPMU22-0002",
    doi = "10.1007/JHEP06(2022)095",
    journal = "JHEP",
    volume = "06",
    pages = "095",
    year = "2022"
}

@article{Geng:2022slq,
    author = "Geng, Hao and Karch, Andreas and Perez-Pardavila, Carlos and Raju, Suvrat and Randall, Lisa and Riojas, Marcos and Shashi, Sanjit",
    title = "{Jackiw-Teitelboim Gravity from the Karch-Randall Braneworld}",
    eprint = "2206.04695",
    archivePrefix = "arXiv",
    primaryClass = "hep-th",
    doi = "10.1103/PhysRevLett.129.231601",
    journal = "Phys. Rev. Lett.",
    volume = "129",
    number = "23",
    pages = "231601",
    year = "2022"
}

@article{Neuenfeld:2024gta,
    author = "Neuenfeld, Dominik and Svesko, Andrew and Sybesma, Watse",
    title = "{Liouville gravity at the end of the world:deformed defects in AdS/BCFT}",
    eprint = "2404.07260",
    archivePrefix = "arXiv",
    primaryClass = "hep-th",
    doi = "10.1007/JHEP07(2024)215",
    journal = "JHEP",
    volume = "07",
    pages = "215",
    year = "2024"
}

@article{Hartle:1983ai,
    author = "Hartle, J. B. and Hawking, S. W.",
    editor = "Fang, Li-Zhi and Ruffini, R.",
    title = "{Wave Function of the Universe}",
    reportNumber = "PRINT-83-0937 (CAMBRIDGE)",
    doi = "10.1103/PhysRevD.28.2960",
    journal = "Phys. Rev. D",
    volume = "28",
    pages = "2960--2975",
    year = "1983"
}

@article{Harlow:2011ke,
    author = "Harlow, Daniel and Stanford, Douglas",
    title = "{Operator Dictionaries and Wave Functions in AdS/CFT and dS/CFT}",
    eprint = "1104.2621",
    archivePrefix = "arXiv",
    primaryClass = "hep-th",
    reportNumber = "SU-ITP-11-22",
    month = "4",
    year = "2011"
}

@article{Hertog:2011ky,
    author = "Hertog, Thomas and Hartle, James",
    title = "{Holographic No-Boundary Measure}",
    eprint = "1111.6090",
    archivePrefix = "arXiv",
    primaryClass = "hep-th",
    doi = "10.1007/JHEP05(2012)095",
    journal = "JHEP",
    volume = "05",
    pages = "095",
    year = "2012"
}

@article{Hayward:1993my,
    author = "Hayward, G.",
    title = "{Gravitational action for space-times with nonsmooth boundaries}",
    doi = "10.1103/PhysRevD.47.3275",
    journal = "Phys. Rev. D",
    volume = "47",
    pages = "3275--3280",
    year = "1993"
}

@article{Harlow:2011ny,
    author = "Harlow, Daniel and Maltz, Jonathan and Witten, Edward",
    title = "{Analytic Continuation of Liouville Theory}",
    eprint = "1108.4417",
    archivePrefix = "arXiv",
    primaryClass = "hep-th",
    reportNumber = "SU-ITP-11-42",
    doi = "10.1007/JHEP12(2011)071",
    journal = "JHEP",
    volume = "12",
    pages = "071",
    year = "2011"
}

@article{Polyakov:1981rd,
    author = "Polyakov, Alexander M.",
    editor = "Khalatnikov, I. M. and Mineev, V. P.",
    title = "{Quantum Geometry of Bosonic Strings}",
    reportNumber = "Print-81-0351 (LANDAU INST)",
    doi = "10.1016/0370-2693(81)90743-7",
    journal = "Phys. Lett. B",
    volume = "103",
    pages = "207--210",
    year = "1981"
}

@article{Hawking:2000da,
    author = "Hawking, Stephen and Maldacena, Juan Martin and Strominger, Andrew",
    title = "{de Sitter entropy, quantum entanglement and AdS / CFT}",
    eprint = "hep-th/0002145",
    archivePrefix = "arXiv",
    doi = "10.1088/1126-6708/2001/05/001",
    journal = "JHEP",
    volume = "05",
    pages = "001",
    year = "2001"
}

@article{Bachas:2001vj,
    author = "Bachas, C. and de Boer, J. and Dijkgraaf, R. and Ooguri, H.",
    title = "{Permeable conformal walls and holography}",
    eprint = "hep-th/0111210",
    archivePrefix = "arXiv",
    reportNumber = "CALT-68-2361, CITUSC-01-045, ITFA-2001-33, LPTENS-01-42",
    doi = "10.1088/1126-6708/2002/06/027",
    journal = "JHEP",
    volume = "06",
    pages = "027",
    year = "2002"
}

@article{Anous:2022wqh,
    author = "Anous, Tarek and Meineri, Marco and Pelliconi, Pietro and Sonner, Julian",
    title = "{Sailing past the End of the World and discovering the Island}",
    eprint = "2202.11718",
    archivePrefix = "arXiv",
    primaryClass = "hep-th",
    doi = "10.21468/SciPostPhys.13.3.075",
    journal = "SciPost Phys.",
    volume = "13",
    number = "3",
    pages = "075",
    year = "2022"
}

@article{Liu:2024oxg,
    author = "Liu, Yan and Lyu, Hong-Da and Wang, Chuan-Yi",
    title = "{On AdS$_{3}$/ICFT$_{2}$ with a dynamical scalar field located on the brane}",
    eprint = "2403.20102",
    archivePrefix = "arXiv",
    primaryClass = "hep-th",
    doi = "10.1007/JHEP10(2024)001",
    journal = "JHEP",
    volume = "10",
    pages = "001",
    year = "2024"
}

@article{Callebaut:2025thw,
    author = "Callebaut, Nele and Selle, Matteo",
    title = "{Setting T$^{2}$ free for braneworld holography}",
    eprint = "2510.01099",
    archivePrefix = "arXiv",
    primaryClass = "hep-th",
    doi = "10.1007/JHEP02(2026)234",
    journal = "JHEP",
    volume = "02",
    pages = "234",
    year = "2026"
}

@article{Callebaut:2026hso,
    author = "Callebaut, Nele and Selle, Matteo",
    title = "{Brane effective actions and their island rule from $T\overline T$ flows}",
    eprint = "2608.30533",
    archivePrefix = "arXiv",
    primaryClass = "hep-th",
    month = "8",
    year = "2026"
}

\bibliographystyle{utphys}

\end{document}